\documentclass[paper,onecolumn,long,noendfloat,new,nonreprod,noblind]{geophysics}

\usepackage{float} 
\usepackage{graphicx} 
\usepackage{caption} 
\usepackage{mathtools}
\usepackage{bm}
\usepackage{amsmath}
\usepackage{amsfonts}
\usepackage{amssymb}
\usepackage{multirow}
\usepackage{placeins}
\usepackage{dirtytalk}
\usepackage{layouts}
\usepackage[font=small,labelfont=small]{caption}
\usepackage[shortcuts]{extdash}
\renewcommand{\b}{\mathbf}

\newcommand{\dr}{\partial}

\renewcommand{\figdir}{Fig}

\begin{document}

\title{Source footprint elimination in full-waveform inversion by model extension: Application to elastic guided waves recorded by distributed acoustic sensing in unconventional reservoir}

	\author{Milad Bader\footnotemark[1], Robert G. Clapp\footnotemark[1], Kurt T. Nihei\footnotemark[2], and Biondo Biondi\footnotemark[1]}

 \author{Milad Bader\footnotemark[1], Robert G. Clapp\footnotemark[1], Kurt T. Nihei\footnotemark[2], and Biondo Biondi\footnotemark[1]}

    \righthead{Footprint elimination by model extension}

    \address{
 
	\footnotemark[1] Stanford University, Department of Geophysics, 397 Panama Mall, Stanford, CA 94305. E-mail: nmbader@sep.stanford.edu; bob@sep.stanford.edu; biondo@sep.stanford.edu.\\
    \footnotemark[2] Chevron Technology Center, 1500 Louisiana Street, Houston, TX 77002. E-mail: kurttnihei@chevron.com.
 }

	\footnotetext[1]{Stanford University, Department of Geophysics, 397 Panama Mall, Stanford, CA 94305.\\E-mail: nmbader@sep.stanford.edu; bob@sep.stanford.edu; biondo@sep.stanford.edu.}
    \footnotetext[2]{Chevron Technology Center, 1500 Louisiana Street, Houston, TX 77002.\\E-mail: kurttnihei@chevron.com.}


\maketitle

\begin{abstract}
Source footprints represent an inherent problem to full-waveform inversion (FWI). They are caused by the high data sensitivity to the model parameters in the vicinity of the seismic sources and can be exacerbated by source-related errors in the modeling operator. We propose a simple, effective, and efficient method to remove source footprints in FWI when sources are located near or inside the volume of interest while robustly updating the model in their vicinity. The method uses illumination redundancy and extends the model along sources. Each source updates one component of the extended model, and a regularization term ensures that these components are mutually consistent, except for their respective footprints. We illustrate the effectiveness of our method on the elastic inversion of synthetic guided waves. We show its robustness in the presence of source-related errors and its superiority over other well-known approaches, such as illumination compensation by inverse pseudo-Hessian and gradient preconditioning. We apply the method to a field distributed acoustic sensing dataset with elastic guided waves generated by perforation shots in an unconventional shale reservoir. The method is able to retrieve localized reservoir anomalies with higher elastic velocities, indicating possible lower pore pressure or tighter shale regions.
\end{abstract}
\section{Introduction}

In recent years, distributed acoustic sensing (DAS) has emerged as a new technology to measure the seismic strain field with unprecedented spatial and temporal resolutions \cite[]{parker2014distributed,li2022distributed}. This technology has opened the door to acquiring more downhole seismic data at a lower cost and without disrupting production or other operations. The recorded data were predominantly for vertical seismic profiling (VSP) to conduct imaging and reservoir monitoring \cite[]{mateeva2014distributed,daley2016field}. Subsequently, more DAS data were acquired in unconventional reservoirs to analyze pressure fronts, and fracture hits using the low-frequency quasi-static strain field \cite[]{jin2017hydraulic,karrenbach2017hydraulic,jin2019machine}. Beyond that, the main application of DAS in unconventional plays remains for detecting, locating, and characterizing microseismicity \cite[]{karrenbach2019fiber,stork2020application,lellouch2022microseismic,huot2022detection}. DAS-recorded seismic waveforms were also used for geometrical fracture imaging \cite[]{lellouch2020fracture,ma2022fracture,stanek2022fracture} and reservoir properties estimation via dispersion analysis \cite[]{luo2021seismic}. These developments paved the way for more advanced geophysical methods, such as full-waveform inversion.

Full-waveform inversion (FWI) is a powerful technique that allows building high-resolution subsurface models by fitting the entire recorded waveforms \cite[]{lailly1983seismic,tarantola1984inversion,mora1987nonlinear,pratt1999seismic,virieux2009overview}. It has become a routine method to build acoustic velocity models in marine environments \cite[]{warner2013anisotropic,prieux2013multiparameter1}. It has also been successfully applied to multicomponent and land datasets \cite[]{prieux2013multiparameter2,vigh2014elastic,raknes2015three,leblanc2022elastic,sedova2022elastic}, and vertical seismic profiling \cite[]{amini2015vertical,owusu2016anisotropic,liu2021elastic}. More recently, DAS-based FWI made its way through VSP acquisitions \cite[]{egorov2018elastic,eaid2020multiparameter}, but its application to DAS data in unconventional reservoirs with \textit{in situ} sources is impeded by a major challenge, that is source footprints. 

Source footprints present an inherent problem to FWI. They are mainly caused by the strong wavefield amplitudes in the vicinity of the seismic sources (and virtual or adjoint sources), leading to a high data sensitivity to the model parameters and producing a high amplitude gradient near the sources. They become particularly pronounced in the case of elastic waves due to the complex source mechanism and near-field effects. Figure \ref{fig:ch5sy_kernels} shows the $V_{\mathrm{P}}$ and $V_{\mathrm{S}}$ sensitivity kernels in 2D homogeneous acoustic and elastic media. The amplitudes in each kernel are clipped symmetrically around zero at 50\% of their maximum absolute value. The sensitivity is the highest near the sources. While the acoustic kernel is relatively smooth, the elastic ones present strong singularities near the sources, which dominate the overall kernels, along with sharp polarity reversals.

\plot[!h]{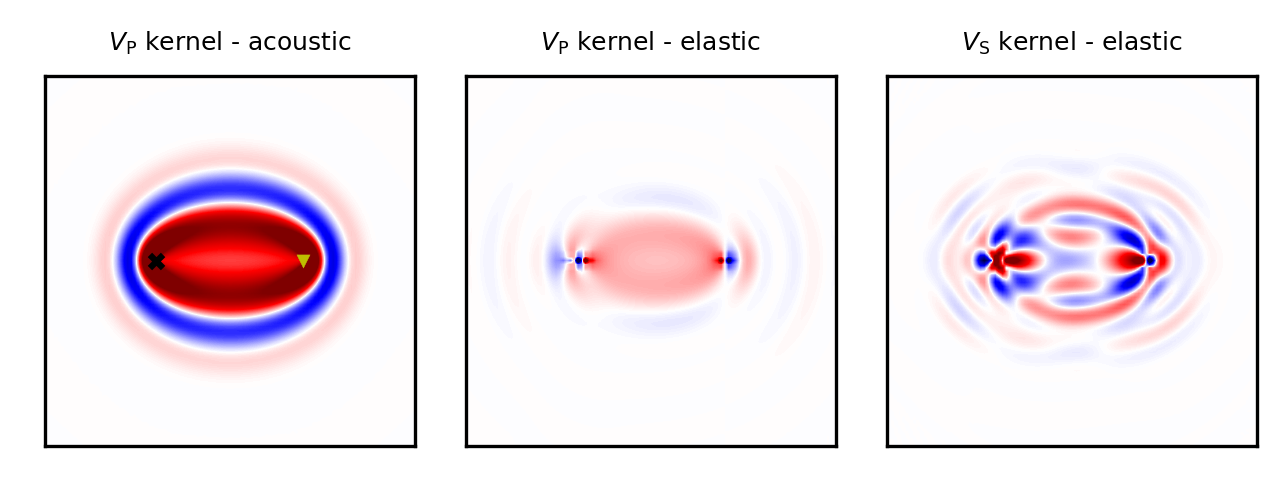}{width=0.95\textwidth}{Examples of FWI misfit sensitivity kernels. The source and adjoint source are indicated by a cross and a triangle, respectively. The source in the elastic case is a horizontal dipole and the amplitudes of each kernel are clipped to 50\% of the maximum absolute value. Notice the strong singularities in $V_{\mathrm{P}}$ elastic kernel and the complicated near source sensitivity in the $V_{\mathrm{S}}$ kernel.}

The footprints problem is well known in seismic imaging and is often mitigated by an operator compensating for the illumination imbalance \cite[]{shin2001improved}. In FWI, the inverse Hessian (Newton method), approximate Hessian (Gauss-Newton method), and pseudo-Hessian (diagonal of the approximate Hessian) account for the geometric amplitude effects \cite[]{virieux2009overview} and should compensate for the source-receiver footprints (Figure \ref{fig:ch5sy_kernels}). In practice, the Hessian and approximate Hessian are seldom constructed (let alone their inverses) due to their prohibitive computation and memory cost. Instead, the application of their inverses to a given vector is computed using iterative methods requiring matrix-vector products, such as linear conjugate-gradient (CG). The matrix-vector product can then be computed using the second-order adjoint state method \cite[]{fichtner2011hessian,metivier2012second}, but the number of iterations is often limited (truncated) so that only an approximate inverse is estimated. In the case of the pseudo-Hessian, it can be efficiently approximated by the zero-lag auto-correlation in time of the forward wavefield or its derivatives \cite[]{shin2001improved}. However, it remains insufficient for multiparameter FWI since it contains no information about the Hessian off-diagonal terms. In contrast, the widely-used limited-memory Broyden–Fletcher–Goldfarb–Shanno method (\textit{l}-BFGS) is a quasi-Newton method comparable in effectiveness with truncated Gauss-Newton and Newton methods at a cost comparable with non-linear CG \cite[]{metivier2013full,operto2013guided,abdul2022gauss}. The method approximates the inverse Hessian at each iteration using model and gradient information from a limited number of previous iterations \cite[]{nocedal2006numerical}. It starts from an initial guess which can approximate the inverse pseudo-Hessian \cite[]{brossier2009seismic}, thus better compensating for source footprints. In all cases, the exact inverse Hessian is not available in practice, and source footprints will not be fully removed. Moreover, when source-related errors in the modeling operator exacerbate these footprints, even an exact Hessian is insufficient. In this case, a possible solution is to mask or taper the FWI gradient by brute force or utilize a regularization to smooth out the footprints. However, these approaches cannot be used when the sources are close or inside the target region, and source-related errors could be important such as perforation shots and microseismic events near or inside unconventional reservoirs.

We propose a simple, effective, and efficient method to remove source footprints in FWI. We use the source illumination redundancy to extend the model and the misfit function along sources. Through a well-designed regularization term, we prevent each source from contributing to the inversion in its vicinity while allowing other potential sources to contribute instead. We demonstrate the effectiveness of our method on synthetic guided-wave inversion examples and show its robustness in the presence of source-related modeling errors. We use our method to estimate $V_{\mathrm{P}}$ and $V_{\mathrm{S}}$ in an unconventional reservoir using \textit{in situ} DAS fiber and perforation shots and highlight lateral heterogeneities that could indicate lower pore pressure in the reservoir.

\section{Theory and Method}

In this section, we give a brief overview of the forward and inverse problems before detailing our method for source footprint attenuation in FWI.

\subsection{Forward problem}
We start by writing the time-domain acoustic or elastic wave equation (WE) in the following compact form
\begin{equation}
    \label{we-general}
    \b{A}(\b{m}(\b{x})) \, \b{u}_s(t, \b{x}) = \b{f}_s(t,\b{x}),
\end{equation}
where $\b{A}$ incorporates the temporal and spatial differential operators and is function of the model $\b{m}$ which encapsulates the acoustic or elastic parameters (P- and S-wave velocities, density, etc.), $\b{u}_s$ is the seismic wavefield (pressure, particle displacement, etc.) induced by the source term $\b{f}_s$, $t$ is time, and $\b{x}=(x,y,z)^T$ is the spatial coordinates vector ($T$ stands for transpose). We consider a synchronous point source which can be written in the case of second-order WEs as
\begin{equation}
    \label{point-source}
    \b{f}_s(t,\b{x}) = 
    \begin{cases}
        & \delta(\b{x}-\b{x}_s)w(t), \quad \text{acoustic pressure source}, \\
        & (f_x,f_y,f_z)^T \delta(\b{x}-\b{x}_s) w(t), \quad \text{elastic vector force}, \\
        & - \b{M} \nabla \delta(\b{x}-\b{x}_s) w(t), \quad \text{moment tensor $\b{M}$ source \allowdisplaybreaks \cite[]{aki2002quantitative}},
    \end{cases}
\end{equation}
where $\delta$ is the Dirac delta function, $\b{x}_s$ is the point source coordinates vector, and $w$ is the source time function. Throughout this paper, we will omit the variables $t$ and $\b{x}$ for conciseness unless it is indispensable. The modeled seismic data vector (or trace) for a given source-receiver pair $s,r$ reads
\begin{equation}
    \label{modeled-data}
    \b{d}_{s,r} = \b{K}_{r} \b{u}_s,
\end{equation}
where $\b{K}_{r}$ is a spatio-temporal sampling operator depending on the receiver location $\b{x}_r$ and on the temporal sampling rate of the wavefield and data. The seismic data depends non-linearily on the model $\b{m}$ via wave equation \ref{we-general}, thus we write $\b{d}_{s,r} = \b{d}_{s,r} (\b{m})$.

\subsection{Solving the inverse problem}
In its simplest form, FWI aims at estimating the model $\b{m}$ by minimizing the misfit, in a least-squares sense, between the observed and synthetic (predicted) data \cite[]{virieux2009overview}. We write this classical misfit function as
\begin{equation}
    \label{misfit-conventional}
    \phi(\b{m}) = \frac{1}{2} \sum_{s=1}^{N_s}\sum_{r\in \mathcal{N}_r^s } \Big\Vert \b{d}_{s,r}(\b{m}) - \b{d}^{obs}_{s,r} \Big\Vert ^2,
\end{equation}
where $N_s$ is the number of sources, $\mathcal{N}_r^s$ is the set of receivers corresponding to the source $s$,  and $\b{d}^{obs}_{s,r}$ is the observed (recorded) trace(s) for a given source-receiver pair.
In practice, misfit \ref{misfit-conventional} is minimized iteratively using a local optimization line search method. At each iteration $k+1$, the update rule can be written in the following general form
\begin{equation}
    \label{update-rule}
    \b{m}_{k+1} = \b{m}_{k} + \alpha_k \b{q}_k,
\end{equation}
where $\alpha_k$ is the step length (positive scalar), and $\b{q}_k$ is the search direction which is required to be a descent direction by most line search algorithms. Typically, $\b{q}_k$ needs at least the computation of the gradient $\b{g}_k$ of misfit \ref{misfit-conventional} with respect to the model $\b{m}_k$ at iteration $k$. The gradient $\b{g}_k$ can be written as
\begin{gather}
    \label{gradient-conventional}
    \begin{split}
        \b{g}_k &= \Big( \frac{\dr \phi}{\dr \b{m}}\Big)_{\b{m}=\b{m}_k}= \sum_{s=1}^{N_s}\sum_{r\in \mathcal{N}_r^s } \Big(\frac{\dr \b{d}_{s,r}}{\dr \b{m}}\Big)_{\b{m}=\b{m}_k}^T \big(\b{d}_{s,r}(\b{m}_k) - \b{d}^{obs}_{s,r}\big) \\
        & = \sum_{s=1}^{N_s} \Big(\frac{\dr \b{d}_{s}}{\dr \b{m}}\Big)_{\b{m}=\b{m}_k}^T \big(\b{d}_{s}(\b{m}_k) - \b{d}^{obs}_{s}\big) = \sum_{s=1}^{N_s} \b{J}_{s,k}^T \big(\b{d}_{s}(\b{m}_k) - \b{d}^{obs}_{s}\big) = \sum_{s=1}^{N_s} \b{g}_{s,k},
    \end{split}
\end{gather}
where we vertically concatenate into a single trace $\b{d}_s$ all traces corresponding to a given source $s$. $\b{J}_{s,k}$ is the data sensitivity matrix for the source $s$ at iteration $k$. It also represents the first-order Born forward modeling operator. Gradient \ref{gradient-conventional} is efficiently computed using the adjoint-state method \cite[]{plessix2006review}, without forming the matrix $\b{J}_{s,k}$ explicitly, by cross-correlating in time the forward and adjoint wavefields (or their derivatives). The exact expressions for $\b{g}_k$ in acoustic and elastic media can be found in \cite{tarantola1984inversion}, \cite{mora1987nonlinear}, \cite{tromp2005seismic}, and \cite{fichtner2010full}.

Setting $\b{q}_k = -\b{g}_k$ corresponds to the steepest descent method (SDM). This method can be ineffective or requires many iterations to converge to an acceptable model \cite[]{pratt1998gauss}. A more popular search direction is the one given by the non-linear conjugate gradient (NLCG) method \cite[]{fletcher1964function}, which linearily combines the gradient $\b{g}_k$ and the previous search direction $\b{q}_{k-1}$. Alternatively, the gradient can be preconditioned such as
\begin{equation}
    \label{gradient-precon}
    \b{q}_{k} = - \b{B}_k \b{g}_k,
\end{equation}
where $\b{B}_k$ is a symmetric non-negative matrix. Among common preconditioners, $\b{B}_k$ can be the inverse of the Hessian $\b{H}_k=\Big( \frac{\dr^2 \phi}{\dr \b{m}^2}\Big)_{\b{m}=\b{m}_k}$ (Newton method), the approximate Hessian $\b{H}_{a,k} = \b{J}_k^T\b{J}_k$ (Gauss-Newton method), the pseudo-Hessian (diagonal of $\b{H}_{a,k}$), or a low-rank approximation of the inverse Hessian (quasi-Newton method such as \textit{l}-BFGS). Here, $\b{J}_k$ is the vertical concatenation of the $N_s$ matrices $\b{J}_{s,k}$.

\subsection{Model extension along sources}

Consider an acquisition configuration where some of the seismic sources are located close or inside the area of interest where FWI is ought to estimate the model parameters. We assume that the region of the model surrounding any such source $s$ can be illuminated by at least one other source located farther away. To prevent the source $s$ from polluting the model by its footprints, we limit its contribution to its surroundings during the inversion and compensate for it by over-expressing the contribution from other relevant sources. Throughout this section, we will designate by "useful model" any inverted model excluding source footprints.

We formalize the above idea by first extending the model $\b{m}$ along sources. The extended model thus reads $\widetilde{\b{m}}=(\b{m}_1,\dots,\b{m}_{N_s})^T$. Then, we extend and regularize the misfit function and write it as
\begin{equation}\label{misfit-extended}
    \phi_{ext}(\widetilde{\b{m}}) = \frac{1}{2} \sum_{s=1}^{N_s}\sum_{r\in \mathcal{N}_r^s } \Big\Vert \b{d}_{s,r}(\b{m}_s) - \b{d}^{obs}_{s,r} \Big\Vert ^2 + \frac{1}{2}\eta^2 \Big\Vert \big(\b{I}-\b{S}^T\b{S}_w\big)\b{W}\widetilde{\b{m}} \Big\Vert^2,
\end{equation}
where the first term is similar to the conventional misfit \ref{misfit-conventional} except for the model $\b{m}_s$ which is now specific to each source $s$, the second term is a regularization term which ensures similarity between the useful models corresponding to different sources, and $\eta$ is a scalar. The operator $\b{W}$ is a square (non-symmetric) matrix composed of diagonal matrices $\b{W}_{ij},\, i,j=1\dots N_s$. For each model $\b{m}_s$, the block-row matrices $\b{W}_{sj}$ replace the part surrounding the source location with a linear combination of all the models $\b{m}_i,\, i=1\dots N_s$ with appropriate weights that we detail later. Therefore, if each model $\b{m}_s$ contains the footprints of its corresponding source only, the output of $\b{W}\widetilde{\b{m}}$ is a set of useful models. Figure \ref{fig:ch5sy_weighting_matrix_2} illustrates the structure of the operator $\b{W}$ in the case of a 1D model extended along three sources. The arrows correspond to the location of the sources. The sum of entries in each row is equal to one. $\b{S}_w$ is a weighted sum operator that collapses the useful models into a single physical (non-extended) one. $\b{S}^T$ is a spreading operator that extends by copying along sources a given physical model. $\b{I}$ is the identity operator. Note that $\b{S}^T$ is not the adjoint of $\b{S}_w$. The latter operator takes into account that different sources illuminate different regions of the model based on their location and the location of the corresponding receivers. Upon minimization of the extended misfit \ref{misfit-extended}, we obtain an optimal extended model $\widehat{\widetilde{\b{m}}}$ where each component $\widehat{\b{m}}_s$ contains the footprints of its corresponding source but shares a similar useful model with the other components. We construct the final physical model as 
\begin{equation}\label{model-final}
    \widehat{\b{m}}=\b{S}_w\b{W}\widehat{\widetilde{\b{m}}},
\end{equation}
which is mainly free of source footprints.

\plot[!h]{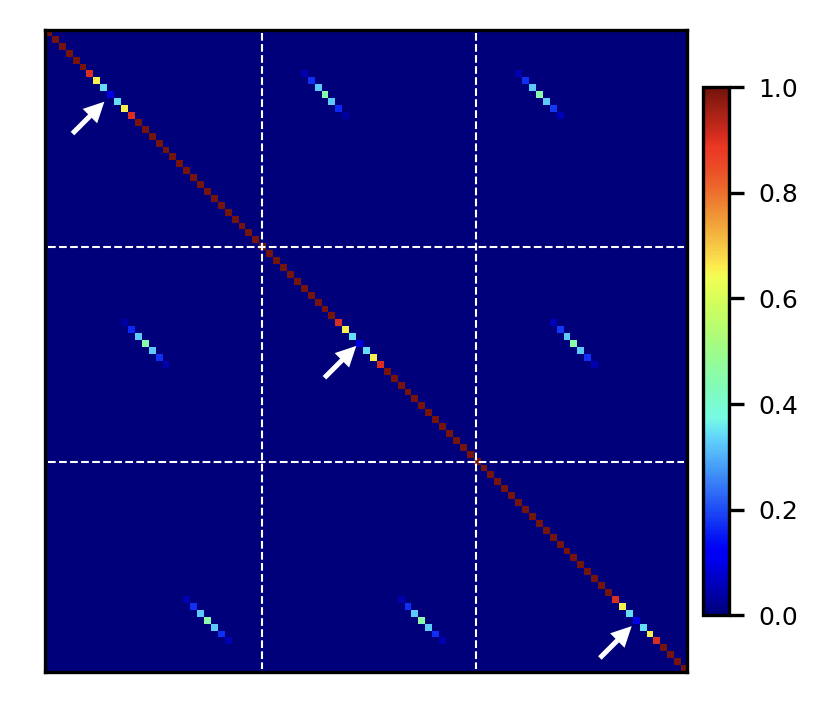}{width=0.95\textwidth}{A realization of the operator (matrix) $\b{W}$ that removes the source footprints from an extended model. The physical model in this case is 1D with 31 samples and is extended along three sources. Each row in the matrix sums to one. The arrows correspond to the location of the sources.}

\subsection{Source zone of influence}

In order to determine the weights used in the operators $\b{W}$ and $\b{S}_w$, we define a zone of influence (ZOI) for each source. The ZOI is described by a weighting vector $\b{b}_s$ with the same spatial dimensions as the physical model. It determines how much a source $s$ should contribute to the model update during the inversion. It can be thought of as a very crude approximation of the source-receiver spread illumination except for the region surrounding the source where the illumination is replaced by its reciprocal. Figure \ref{fig:ch5sy_setting}a shows a three-layer $V_{\mathrm{P}}$ model where the middle layer of interest is a waveguide. We place sources (crosses) and receivers (dots) inside the waveguide to mimic the configuration of an unconventional reservoir crossed by a horizontal well equipped with a DAS fiber. Taking the central source as an example, Figure \ref{fig:ch5sy_setting}b shows the corresponding $\b{b}_s$ vector. The weights smoothly decay from 1 to a small positive value at the source location (forming a "sinkhole"), and away from the source-receiver spread in the horizontal direction. Note that we can also limit the ZOI in the vertical direction but it is unnecessary here since we are interested in the waveguide only. For a more complicated source-receiver geometry, an estimate of the illumination can be used to construct the ZOI (except around the source). However, this is unnecessary since the illumination compensation can be accounted for by the data term inverse Hessian or its approximation. In our implementation, we use the following weights for the ZOI transition at the outer boundaries and in the source sinkhole
\begin{equation}\label{zoi}
    b_s= \frac{\cos^n(\frac{1}{2}\pi r_s) - \log(\nu)}{1-\log({\nu})},
\end{equation}
where $0\leq r_s \leq 1$ is a normalized distance away from the active part of the ZOI (where $b_s=1$), $n$ and $\nu$ ($0< \nu < 1$) determine the stiffness and height of the ZOI transition, respectively. We use the Euclidean distance for the outer taper and Mahalanobis distance around the source location to allow for an ellipsoidal sinkhole. Thus, for each model $\b{m}_s$, the operator $\b{W}$ replaces the values located inside the source sinkhole by
\begin{equation}\label{weights}
    m_s \leftarrow b_s m_s + \frac{1-b_s}{\sum_{i=1,i\neq s}^{N_s}{b_i}} \sum_{i=1,i\neq s}^{N_s}{b_i m_i},
\end{equation}
and leaves the rest of the locations unchanged. As for the operator $\b{S}_w$, it performs the weighted sum
\begin{equation}\label{weighted-sum}
    m = \frac{\sum_{s=1}^{N_s}{b'_s m_s}}{\sum_{s=1}^{N_s}{b'_s}},
\end{equation}
where $b'_s = b_s$ outside the source sinkhole and $b'_s = 1$ inside. There is no need to account for the source footprints in the operator $\b{S}_w$ as they are handled by the operator $\b{W}$.

\plot[!h]{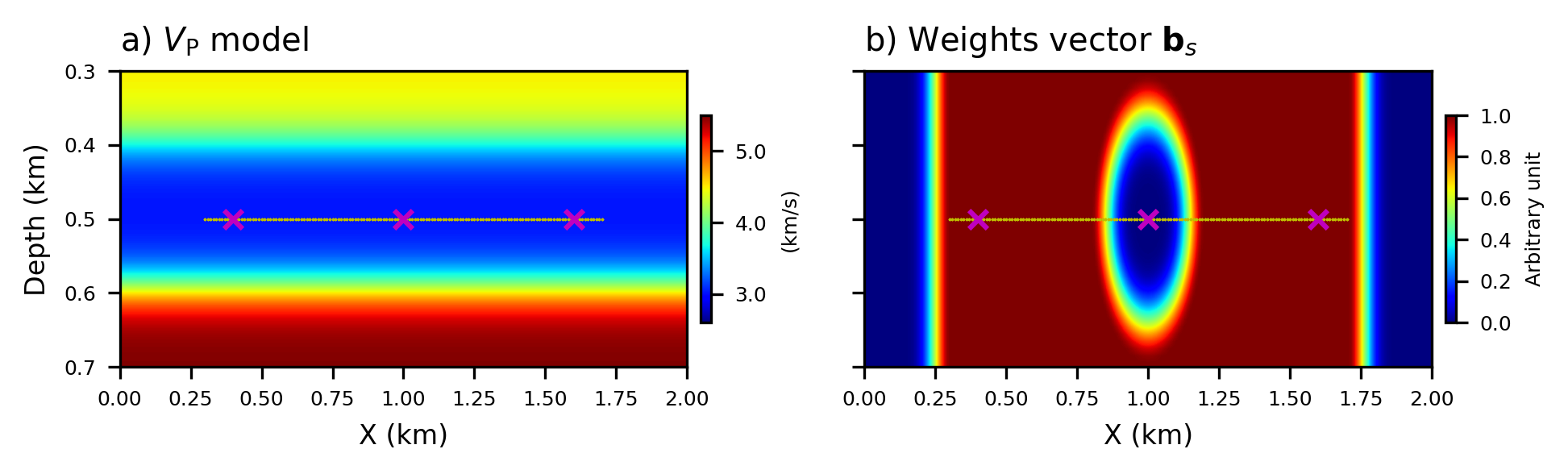}{width=0.95\textwidth}{a) Synthetic three-layer $V_{\mathrm{P}}$ model with sources and receivers aligned in the middle layer. b) Extension weights vector corresponding to the central source. Note that the aspect ratio of the figure is not 1, and the vertical dimension is stretched compared to the horizontal.}

\section{Synthetic examples}

We consider the 2D model and acquisition configuration presented in Figure \ref{fig:ch5sy_setting}a. The three sources have the same diagonal moment tensor (MT) where its components satisfy $M_{33}/M{_{11}}=0.75$ and a Ricker wavelet time function with a 10 Hz central frequency. The receivers are geophones recording two-components particle velocity. The initial model is elastic isotropic with the parameters of each layer listed in Table \ref{table:elastic-parameters}, and the transition between layers is smooth. We consider the Gaussian perturbations inside the waveguide in Figure \ref{fig:ch5sy_true_perturbations} to be recovered by FWI. We define the perturbation of a given model in terms of percentage with respect to the starting model: $\text{perturbation}=100\times(\text{given model} - \text{starting model})/\text{starting model}$. No density perturbation is considered in this example.

\begin{table}[h!]
\centering
\begin{tabular}{ |c||c|c|c| } 
 \hline
 Layer & $V_{\mathrm{P}}$ (km/s) & $V_{\mathrm{S}}$ (km/s) & Density (g/cc) \\ 
 \hline
 \hline
 Top & 4.5 & 2.3 & 2.3 \\ 
 \hline
 Middle & 3.0 & 1.7 & 2.0 \\ 
 \hline
 Bottom & 5.5 & 2.7 & 2.5 \\ 
 \hline
\end{tabular}
\caption{Elastic parameters of the starting model.}
\label{table:elastic-parameters}
\end{table}

\plot[!h]{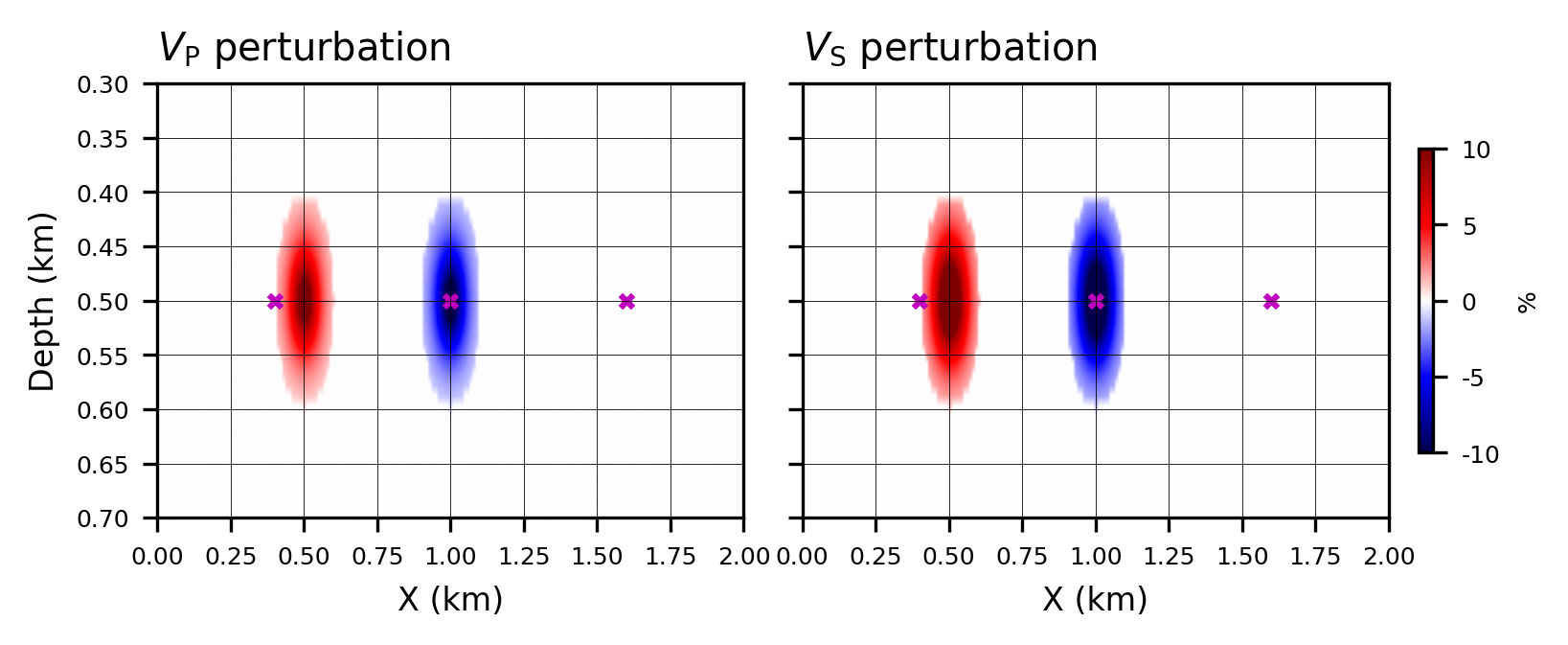}{width=0.95\textwidth}{True $V_{\mathrm{P}}$ and $V_{\mathrm{S}}$ Gaussian perturbations to be recovered by FWI. The structure of the starting model is shown in Figure \ref{fig:ch5sy_setting}a.}

\subsection{Inversion without source errors}

In the first test, we generate noise-free data using the true model and invert them using the exact source parameters (moment tensor, time function, and location) in the conventional (misfit \ref{misfit-conventional}), and extended (misfit \ref{misfit-extended}) FWI cases. The starting model for the inversion excludes the Gaussian perturbations. We use an \textit{l}-BFGS solver with a step length satisfying the Wolfe conditions \cite[]{nocedal2006numerical}. To compare results from different inversions, it is more adequate to set up a convergence criterion rather than fixing the number of iterations. Thus, we define the convergence rate at each iteration $k+1$ as
\begin{equation}
    \label{convergence-rate}
    rate_{k+1} = \frac{\Phi_{k} - \Phi_{k+1}}{\Phi_{k}},
\end{equation}
where $\Phi_{k}$ is any misfit function at iteration $k$, and we stop the inversion when the rate is less than 0.1\% or when the solver fails to find a suitable step. Figure \ref{fig:ch5sy_fwi0}a and \ref{fig:ch5sy_fwi0}b shows the inverted perturbations for the conventional and extended FWI, respectively. The anomalies are easily distinguishable in both cases. However, source footprints are still visible in the conventional FWI whereas the extended FWI was able to remove them. Figure \ref{fig:ch5sy_data_misfit_noerror} shows the normalized data misfit functions obtained from the conventional misfit \ref{misfit-conventional} and the first term in the extended misfit \ref{misfit-extended}. The convergence criterion is reached after 45 iterations in conventional FWI and 116 iterations in extended FWI. At equal number of iterations below 45, the data misfit for extended FWI is higher as expected from a regularized inversion. However, the extended FWI was able to catch up at later iterations and further reduce the data misfit. Note that enforcing more iterations in conventional FWI further reduced the data misfit but failed to remove the remaining source footprints.

\plot[!h]{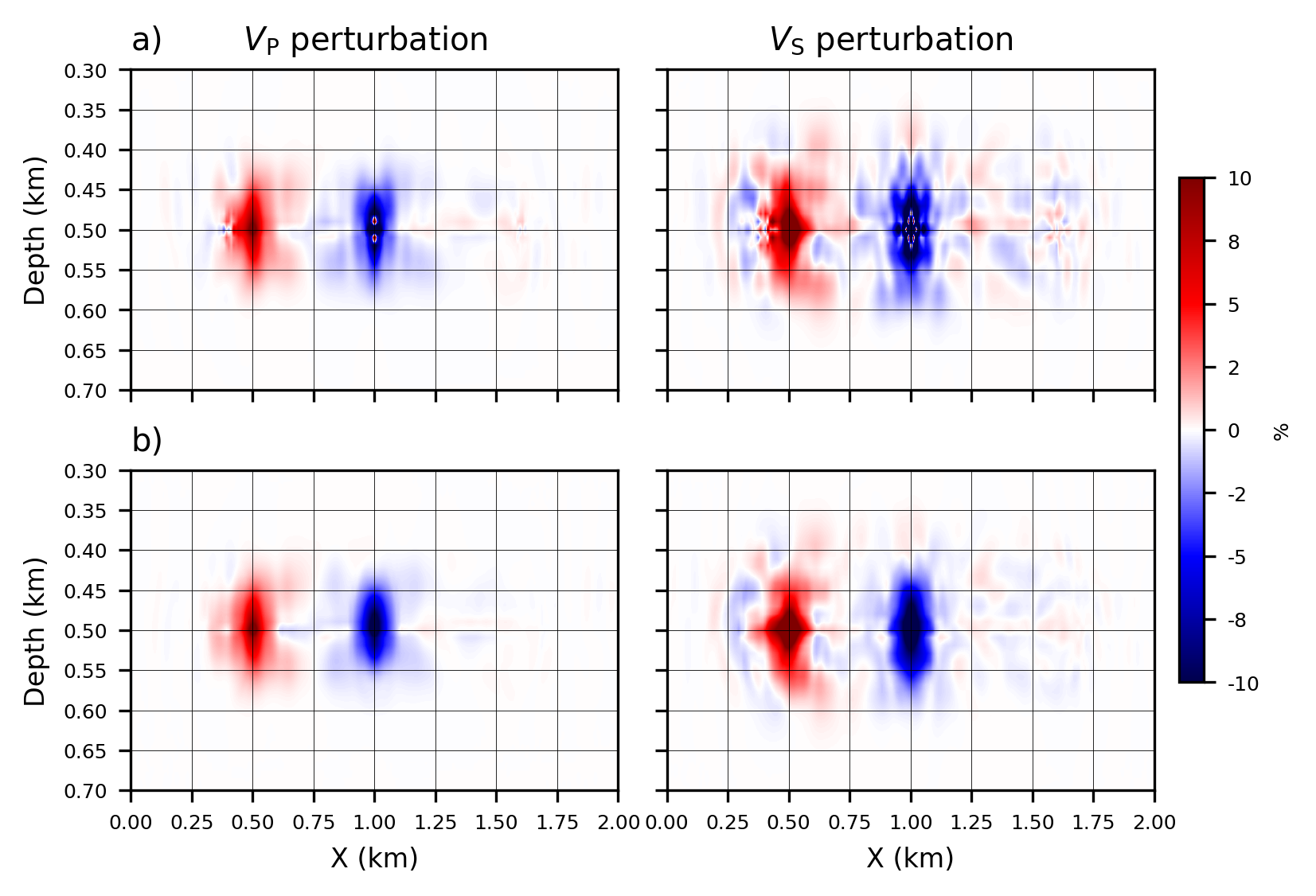}{width=0.95\textwidth}{Inverted perturbations using exact source parameters in a) conventional and b) extended FWI. The Gaussian anomalies are well recovered. Remaining source footprints that are still visible in the conventional FWI were removed by extended FWI.}

\plot[!h]{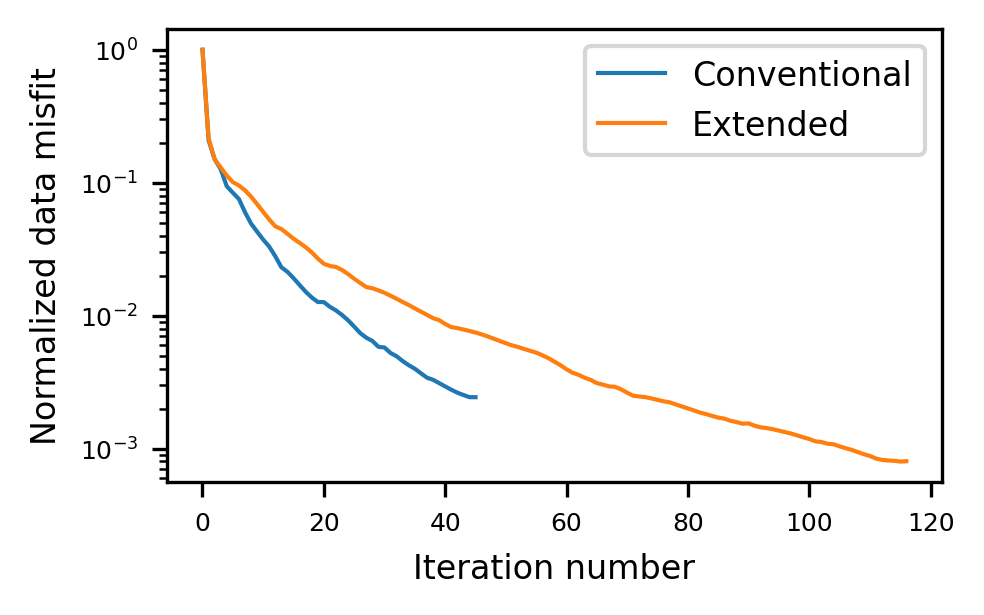}{width=0.95\textwidth}{Normalized data misfit for conventional and extended FWI in the case of exact source parameters.}

We obtain the model perturbation shown in Figure \ref{fig:ch5sy_fwi0}b from the final physical model given by equation \ref{model-final}. To examine the actual inverted model $\widehat{\widetilde{\b{m}}}$ in extended FWI, we show in Figure \ref{fig:ch5sy_fwi0e_ext}a, \ref{fig:ch5sy_fwi0e_ext}b, and \ref{fig:ch5sy_fwi0e_ext}c the extended perturbations for the left-side, central, and right-side sources, respectively. Each extended perturbation is similar to the other two (and to the final model in Figure \ref{fig:ch5sy_fwi0}b) except for the corresponding source footprints.

\plot[!h]{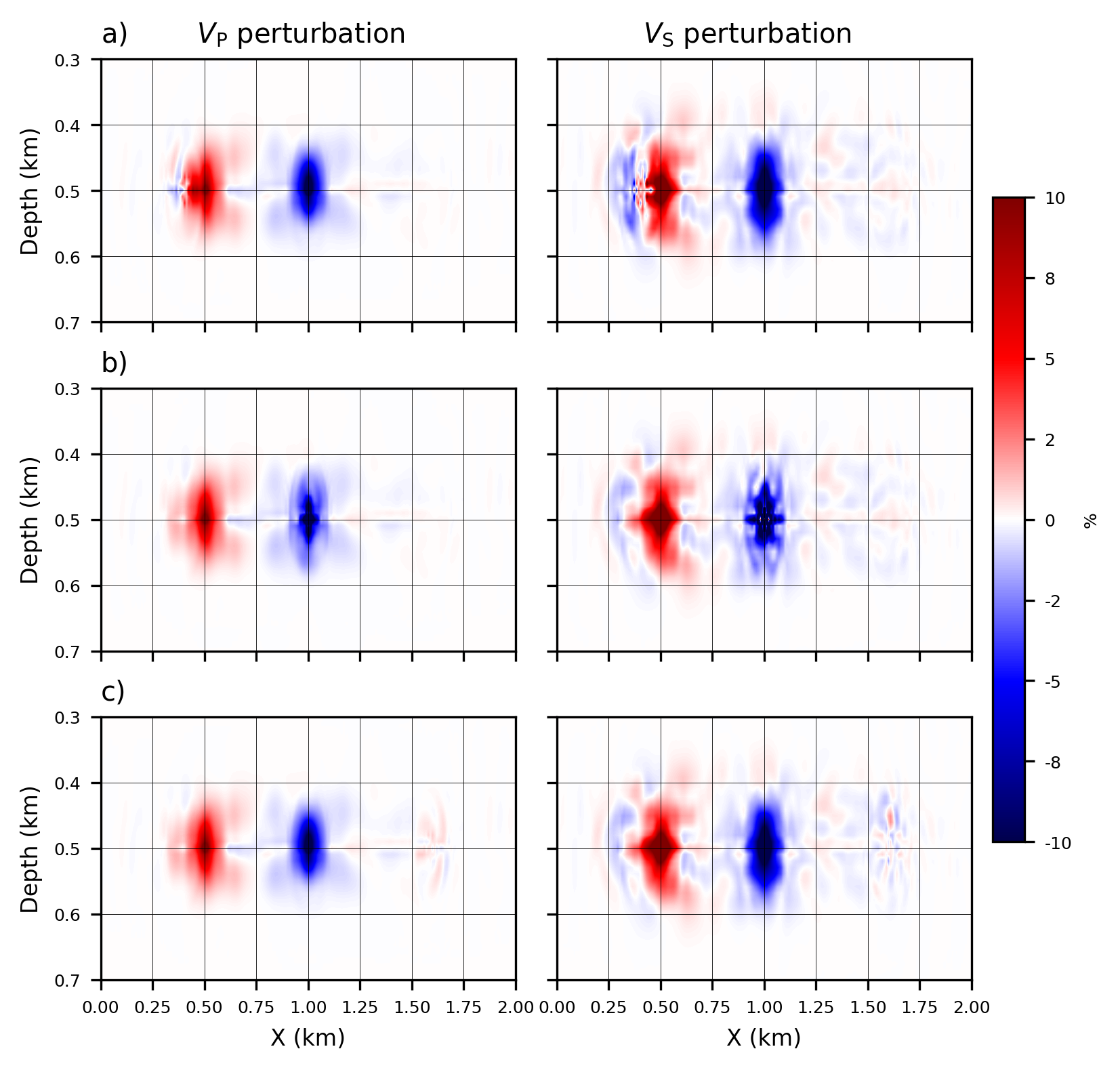}{width=0.95\textwidth}{Inverted perturbations for the a) left-side, b) central, and c) right-side source models in extended FWI.}

\subsection{Inversion with source-related errors}

In the second test, we introduce errors to the source parameters, then we perform the same inversions as in the previous test. We introduce the errors to mimic field applications with perforation shots or microseismic sources. In these applications, the source parameters are typically estimated from the data and suffer inherently from uncertainties and noise \cite[]{bader2023-ea}. Thus, we shift all three sources 1 m to the left. We set the MT ratio to $M_{33}/M_{11}=0.964$ while maintaining its exact Frobenius norm and add an arbitrary 16.5\% relative error to the time function as shown in Figure \ref{fig:ch5sy_wavelet}. Figure \ref{fig:ch5sy_fwi5}a and \ref{fig:ch5sy_fwi5}b shows the inverted perturbations for the conventional and extended FWI, respectively. The source-related errors had a dramatic effect on conventional FWI in this acquisition setting. The true anomalies are no longer recognizable, in particular the $V_{\mathrm{S}}$ anomalies, and fake anomalies are introduced at the location of the right-side source. Conversely, the extended FWI was able to recover the $V_{\mathrm{P}}$ anomalies fairly well, and the $V_{\mathrm{S}}$ anomalies can still be distinguished despite the smearing effect. Table \ref{table:cosine-similarity} summarizes the cosine similarity between the inverted perturbations and the true anomalies for conventional and extended FWI with and without source-related errors. Recall that the cosine similarity between two vectors $\b{v}_1$ and $\b{v}_2$ is given by
\begin{equation}
    cs = \frac{\b{v}_1^T\b{v}_2}{\Vert \b{v}_1\Vert \Vert \b{v}_2 \Vert}, \quad -1 \leq cs \leq 1,
\end{equation}
and it reaches 1 when the two vectors differ only by a scaling factor. It is clear from Table \ref{table:cosine-similarity} that extended FWI recovers perturbations that are structurally consistent with the true anomalies compared to conventional FWI.

\plot[!h]{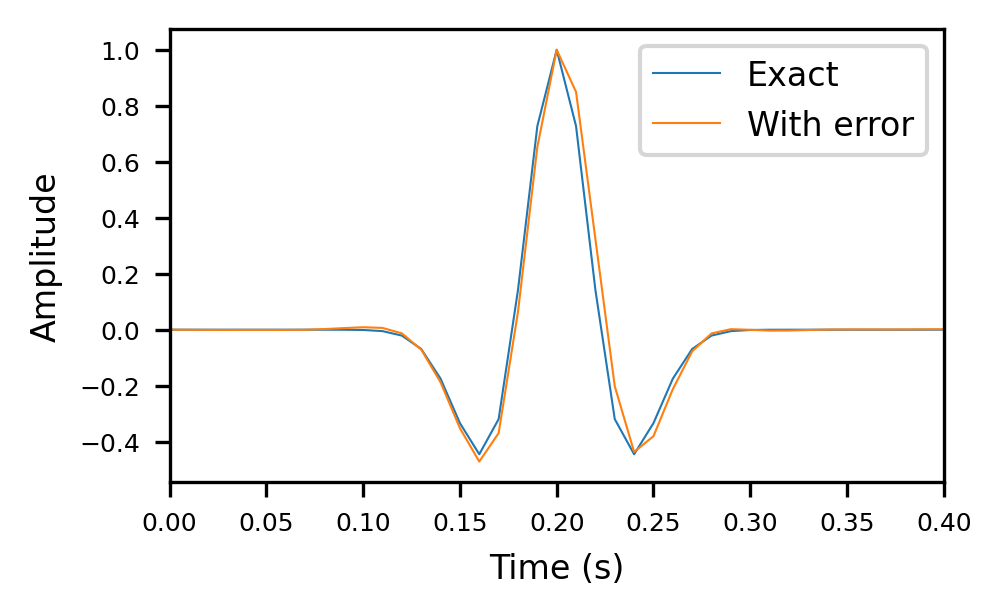}{width=0.95\textwidth}{Source time functions with and without errors used in synthetic FWI tests.}

\plot[!h]{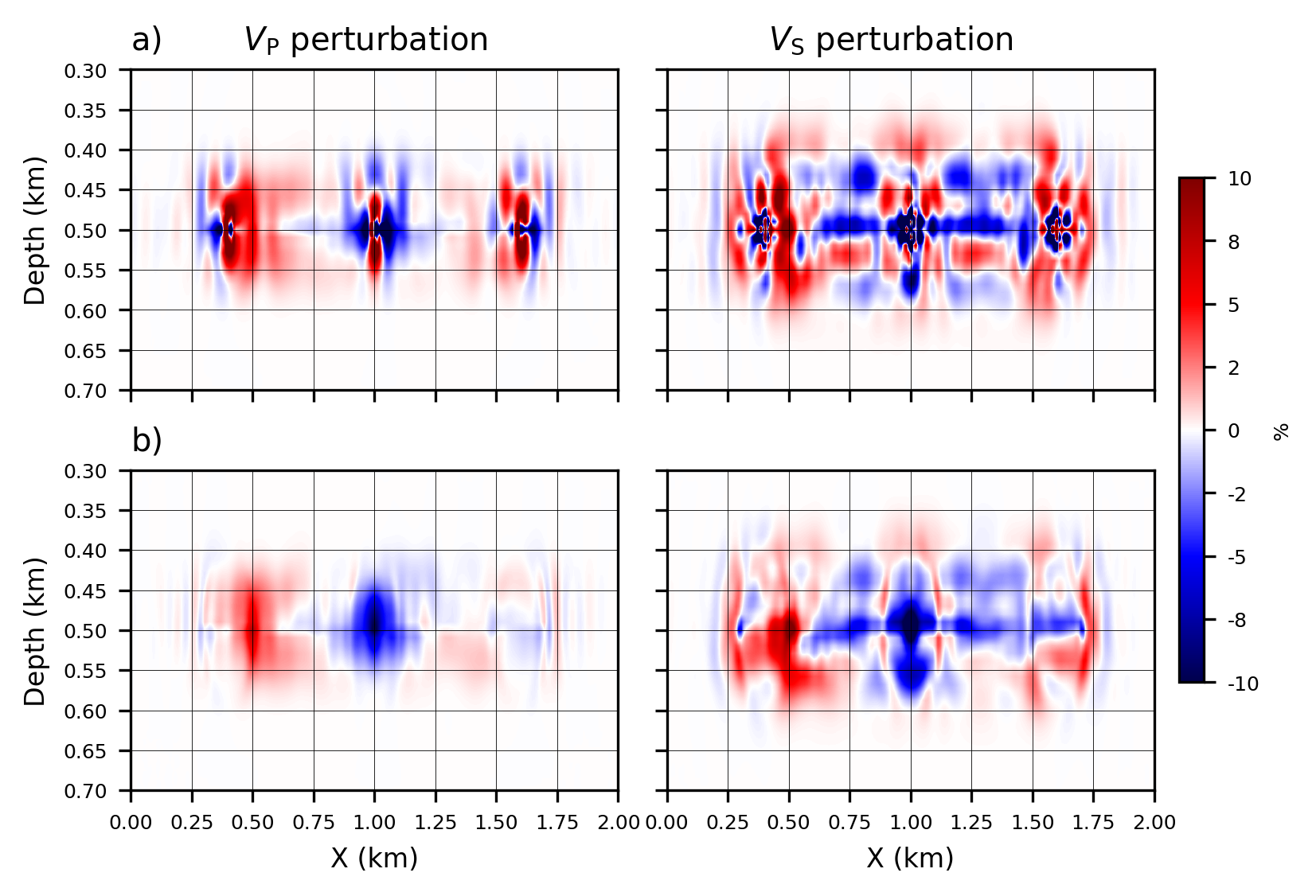}{width=0.95\textwidth}{Inverted perturbations using incorrect source parameters in a) conventional and b) extended FWI. The true Gaussian anomalies can no longer be easily identified in conventional FWI where source footprints are predominant. In extended FWI, the anomalies are well recovered for $\b{V}_{\mathrm{P}}$ and can be structurally identified for $\b{V}_{\mathrm{S}}$. Source footprints have been greatly reduced compared to conventional FWI.}

\begin{table}[h!]
\centering
\resizebox{\columnwidth}{!}{
\begin{tabular}{ |c||c|c|c|c| } 
 \hline
 Parameter & Conventional & Extended & Conventional with errors &  Extended with errors\\ 
 \hline
 \hline
 $V_{\mathrm{P}}$ & 0.83 & 0.96 & 0.19 & 0.89 \\ 
 \hline
 $V_{\mathrm{S}}$ & 0.77 & 0.96 & 0.30 & 0.70 \\ 
 \hline
\end{tabular}
}
\caption{Cosine similarity between inverted perturbations and true anomalies for conventional and extended FWI with and without source-related errors.}
\label{table:cosine-similarity}
\end{table}

We investigated two alternative methods to reduce the source footprints in conventional FWI in the presence of source-related errors. In the first method, we muted out near offsets ($<200$ m) and provided the \textit{l}-BFGS solver with an approximate inverse pseudo-Hessian operator as initial guess for the inverse Hessian at every iteration \cite[]{nocedal2006numerical}. We built this diagonal operator by auto-correlating in time the absolute particle velocity wavefields, summing over shots, clipping, and scaling. Figure \ref{fig:ch5sy_invHessian} shows the constructed operator which scales down the FWI gradient near the sources. In the second method, we applied our source footprint removal operator to the concatenation of individual gradients $\b{g}_{s,k}$ before summing over shots such that
\begin{equation}
    \Bar{\b{g}}_k=\b{S}_w\b{W}\widetilde{\b{g}}_k, \quad \widetilde{\b{g}}_k = (\b{g}_{1,k},\dots,\b{g}_{N_s,k})^T,
\end{equation}
where $\Bar{\b{g}}_k$ is the modified gradient provided to the optimization solver. This type of gradient preconditioning has been used heuristically in regional adjoint waveform tomography to cut out the regions surrounding earthquakes and stations and ensure smooth model updates \cite[]{rodgers2022wus256}. Figure \ref{fig:ch5sy_fwi5_mh_p}a and \ref{fig:ch5sy_fwi5_mh_p}b shows the inverted perturbations using the first and second method, respectively. The first method reduced the source footprints compared to Figure \ref{fig:ch5sy_fwi5}a, but the true anomalies are still difficult to discern. The footprints seem removed or smoothed out by the second method, but the recovered perturbations are clearly far from the true ones. Even though we used the same operator in the second method as in our original extended FWI formulation, altering the FWI gradients before summation over sources is an invalid operation from an optimization point of view. Table \ref{table:cosine-similarity-2} summarizes the cosine similarity between the inverted perturbations and the true anomalies for the two alternative methods. It is clear from the cosine similarity metric that the extended FWI is superior to both methods.

\begin{table}[h!]
\centering
\resizebox{0.6\columnwidth}{!}{
\begin{tabular}{ |c||c|c| } 
 \hline
 Parameter & Mute \& pseudo-Hessian & Gradient preconditioning\\ 
 \hline
 \hline
 $V_{\mathrm{P}}$ & 0.48 & 0.37 \\ 
 \hline
 $V_{\mathrm{S}}$ & 0.30 & 0.19 \\ 
 \hline
\end{tabular}
}
\caption{Cosine similarity between inverted perturbations and true anomalies for FWI with source-related errors using near-offset mute combined with inverse pseudo-Hessian or heuristic gradient preconditioning.}
\label{table:cosine-similarity-2}
\end{table}

We show in Figure \ref{fig:ch5sy_data_misfit_witherror} the normalized data misfit for the two alternative methods and conventional and extended FWI. The relative decrease in data misfit is the greatest for the first method above (mute + pseudo-Hessian), yet the inverted perturbations are not the most accurate. Note that the initial absolute misfit value for this method is different from the others since near offsets were muted. The heuristic gradient preconditioning causes the solver to stop early without finding an appropriate step length, probably because this preconditioning is not consistent with the optimization problem. The extended FWI fits the data better than conventional FWI, although it needs more iterations to reach the same convergence rate. Note that adding a regularization term to the conventional FWI such as second-order Tikhonov \cite[]{aster2018parameter} will have the effect of smearing the footprints instead of removing them.

\plot[!h]{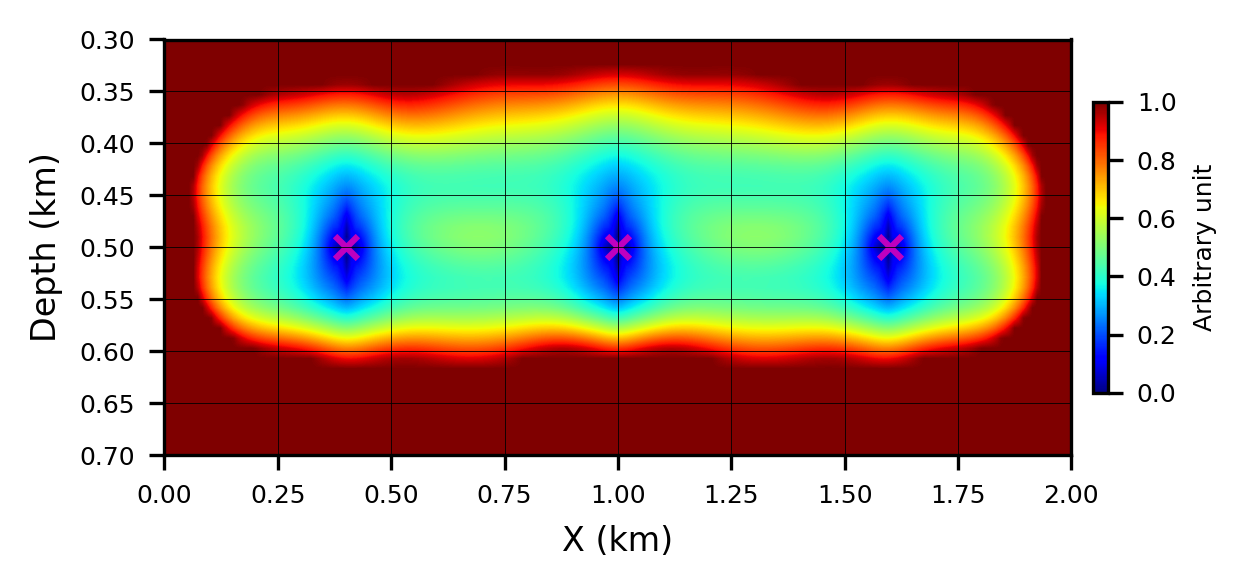}{width=0.95\textwidth}{Approximate inverse pseudo-Hessian operator provided to the (\textit{l}-BFGS) solver at every iteration to precondition the FWI gradient.}

\plot[!h]{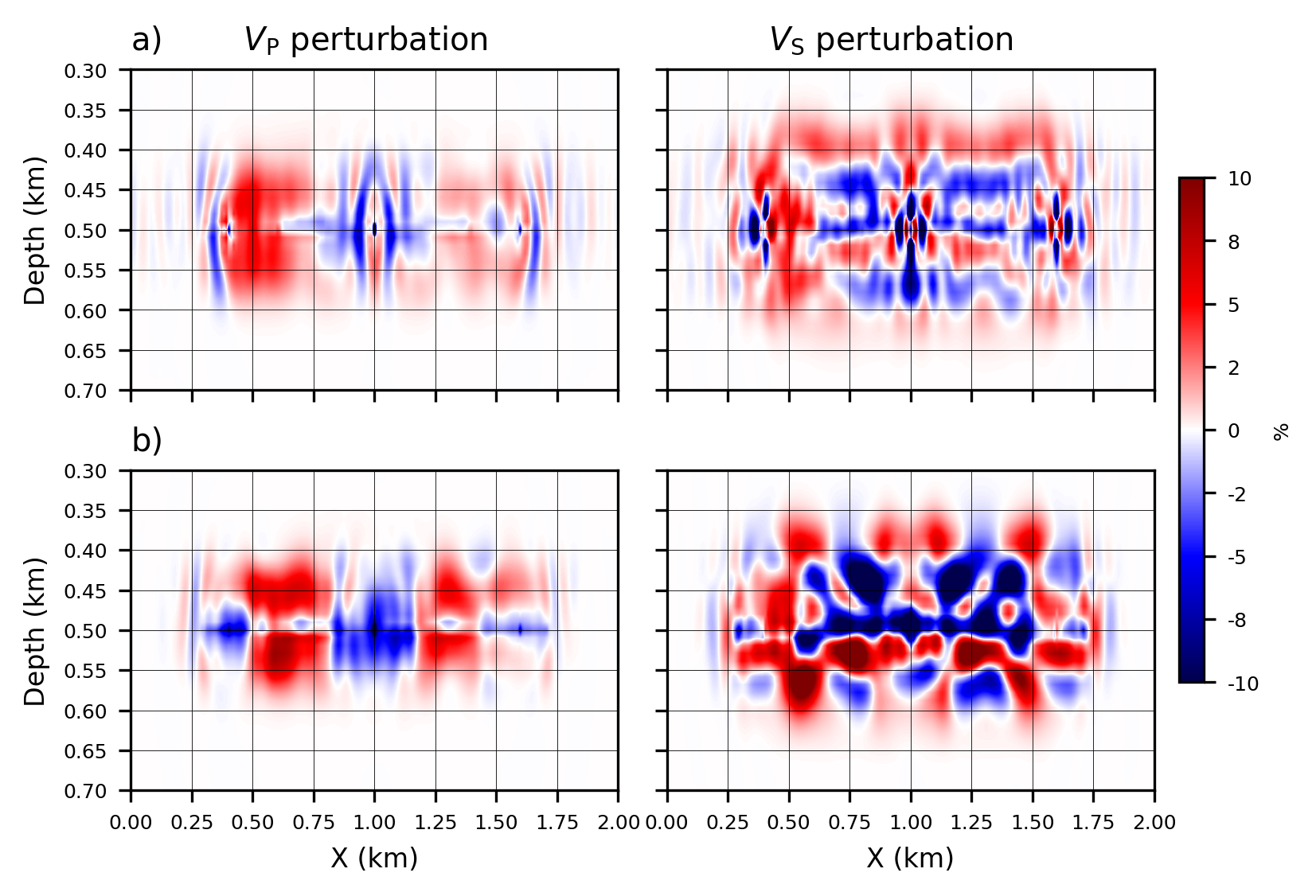}{width=0.95\textwidth}{Inverted perturbations using incorrect source parameters in conventional FWI a) after muting near offsets and providing an approximate inverse pseudo-Hessian operator, and b) with gradient preconditioning before summation over sources. The footprints are attenuated in a) compared to conventional FWI in Figure \ref{fig:ch5sy_fwi5}a but the true anomalies are hardly discernible. In b) the footprints are removed by the preconditioning but the inverted perturbations are incorrect.}

\plot[!h]{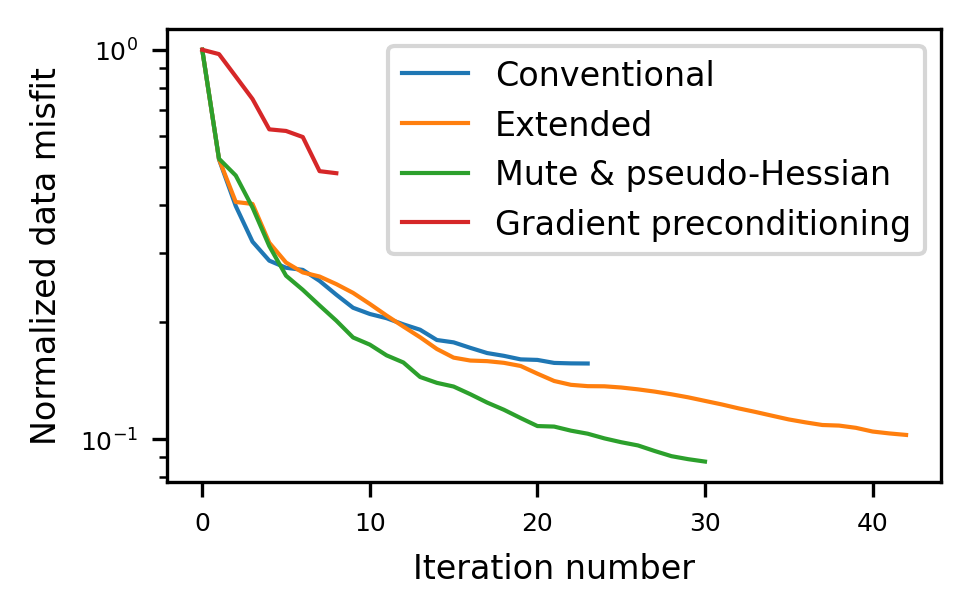}{width=0.95\textwidth}{Normalized data misfit for conventional FWI, extended FWI, FWI with near-offset mute and inverse pseudo-Hessian approximation, and FWI with gradient preconditioning before summation over sources.}

\section{Application to perforation shots and DAS data in unconventional reservoir}

\subsection{Data and inversion setup}

We applied our method to a field data set acquired with a DAS fiber cemented behind a deviated well in an unconventional shale reservoir. The fiber measures strain rate with 10 m gauge length and 1 m channel spacing. The seismic sources are the perforation shots located in the same well and used for hydraulic stimulation. Each source is represented by a diagonal moment tensor as proposed in \cite{bader2023-ea}. We applied a simple processing to the data as described in Bader et al. (2023), which includes muting tube waves, conversion from 3D to 2D \cite[]{auer2013critical}, mild f-k de-noising, resampling to 2 ms, and empirical amplitude-vs-offset (AVO) correction. Figure \ref{fig:ch5un_data} shows a processed DAS shot gather example. The unconventional reservoir layer is a waveguide that traps (quasi) P- and S-waves over hundreds of meters as described in \cite{lellouch2019observations}. We use the seismic arrivals recorded at negative offsets in our inversions. These arrivals have traveled in the region of the reservoir formation that hasn't been stimulated yet (heel side). Hence, they are not affected by the perturbations caused by hydraulic stimulation and carry information about the reservoir formation in its original state.

\plot[!h]{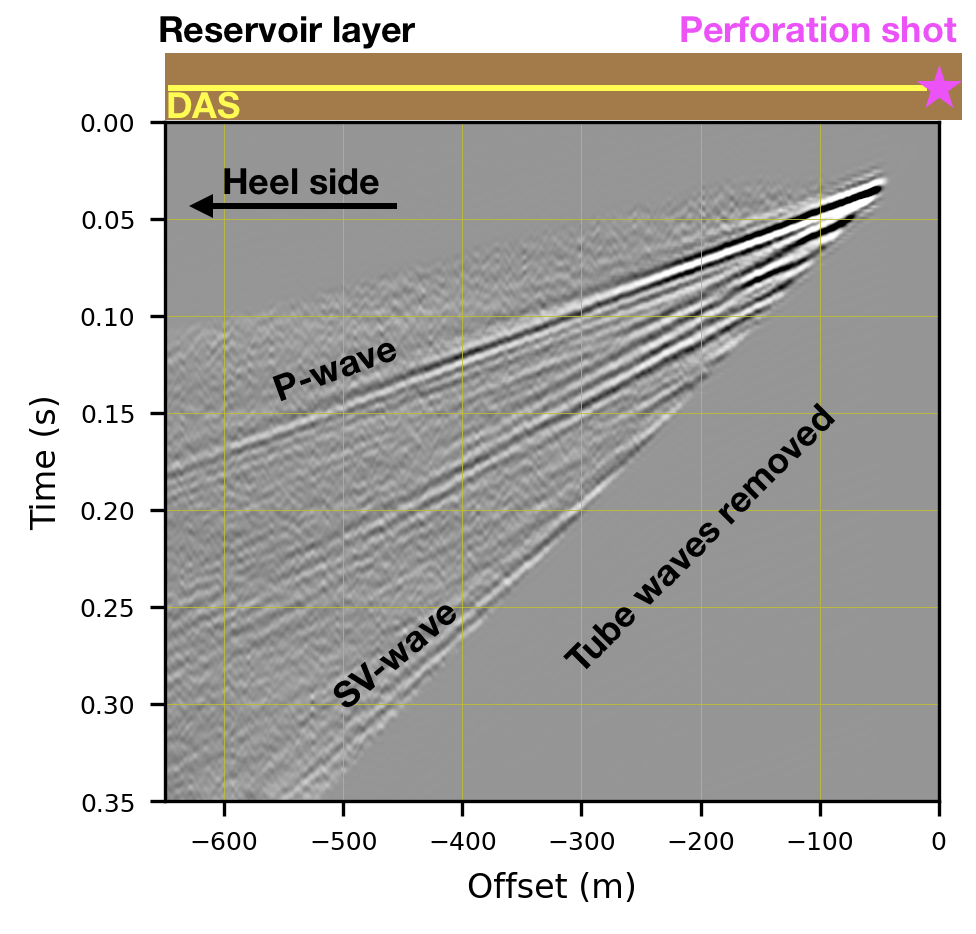}{width=0.95\textwidth}{Processed DAS shot gather example with a sketch on top depicting the acquisition geometry. Reservoir-guided P- and S-waves are recorded by the DAS fiber. We muted out the tube waves and selected negative offsets which illuminate the reservoir layer in its original state before hydraulic stimulation.}

We built our starting 2D VTI model for FWI using logs from a nearby vertical well. We extrapolated the 1D profiles laterally into a structurally conformal model using the formation bottom horizon. Figure \ref{fig:ch5_model_vp} shows the 2D vertical P-wave velocity model ($V_{\mathrm{P}0}$) along with the 16 sources and DAS portion used in the inversions. The dashed line indicates the formation bottom horizon. The model has a grid spacing of 2 and 1 m in the horizontal and vertical directions, respectively. To build the Thomsen anisotropy parameter $\epsilon$ \cite[]{thomsen1986weak}, we scaled the vertical gamma ray profile by the optimal scalar that best matches the P-arrival between synthetic and field data for a selected shot gather. As for the Thomsen anisotropy parameter $\delta$, it cannot be well constrained from the data or well logs. Thereofre, we set it to $\frac{1}{2}\epsilon$. In all inversions, we keep density and anisotropy fixed and update $V_{\mathrm{P}0}$ and $V_{\mathrm{S}0}$ only. We also apply a mask to FWI gradients to restrict the update to the reservoir layer. Figure \ref{fig:ch5un_vertical_profiles} shows the vertical profiles from the starting 2D VTI model and the gradient mask.

\plot[!h]{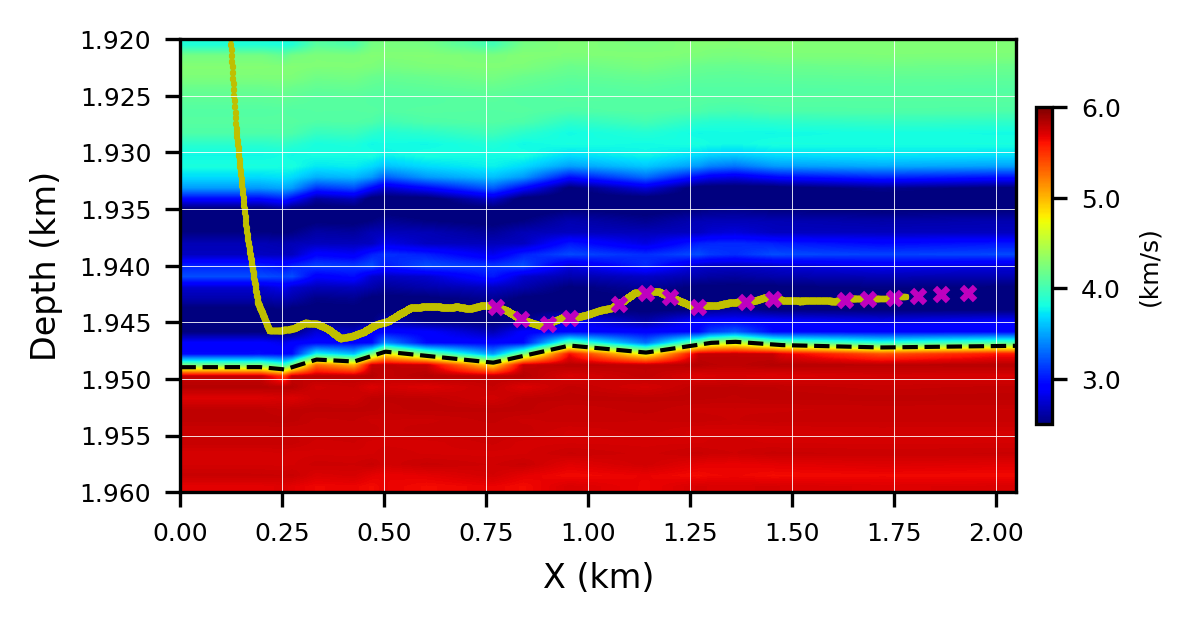}{width=0.95\textwidth}{Vertical P-wave velocity model ($V_{\mathrm{P}0}$) part of the structurally conformal 2D VTI model. The crosses correspond to 16 perforation shots used in FWI and the yellow line indicates the corresponding portion of the DAS fiber. The unconventional reservoir formation is a 15 m thick low-velocity layer and its bottom is indicated by the dashed line.}

\plot[!h]{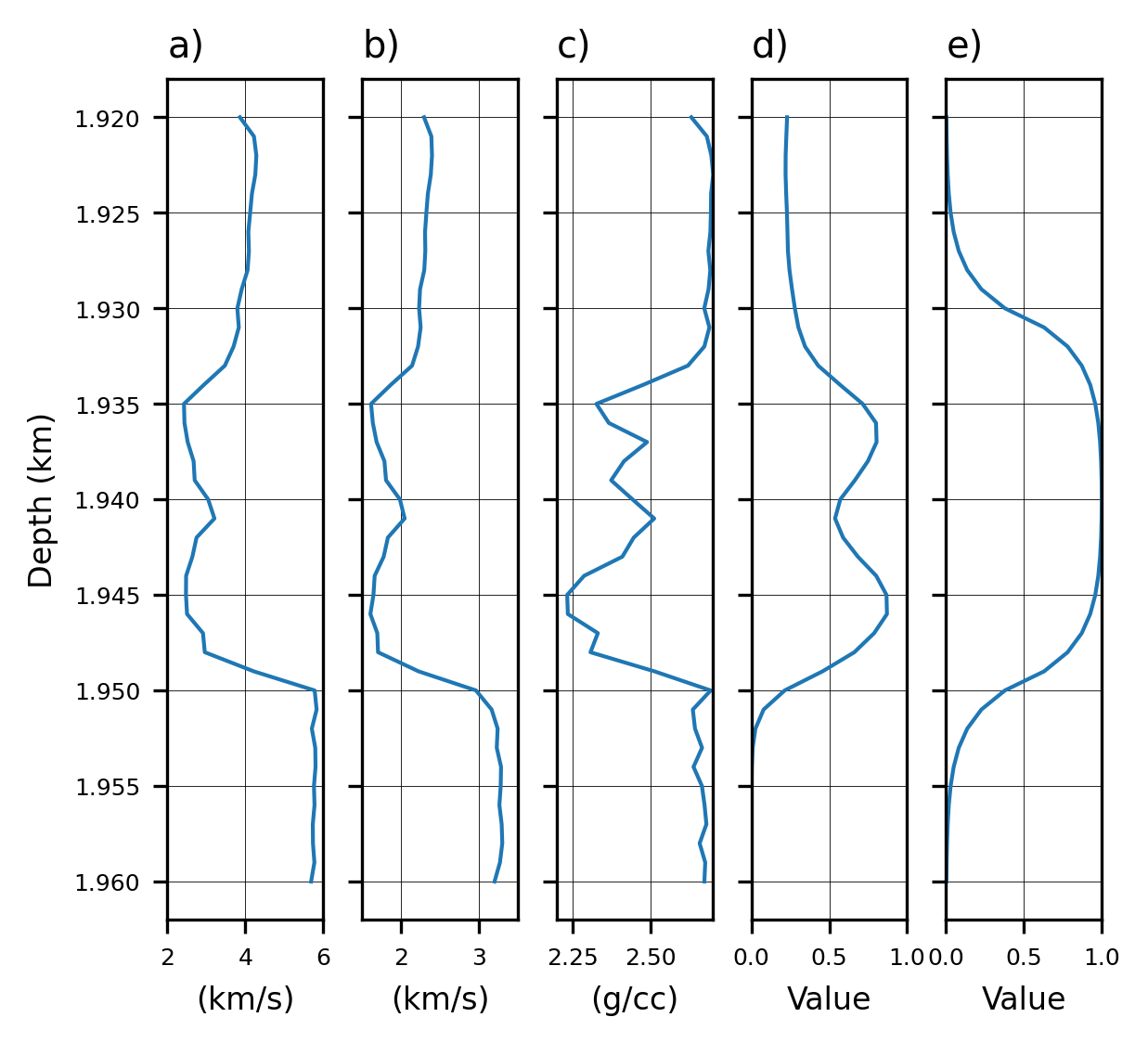}{width=0.95\textwidth}{Vertical profiles from the starting 2D VTI model. a) $V_{\mathrm{P}0}$. b) $V_{\mathrm{S}0}$, c) Density. d) Thomsen anisotropy parameter $\epsilon$. e) FWI gradient mask. The anisotropy parameter $\delta$ is set to $\frac{1}{2}\epsilon$.}

\subsection{FWI results}

We performed conventional and extended FWI using the DAS data in the offset range  [$-650$ - $-200$ m] and the frequency bandwidth [$30$ - $100$ Hz] with a peak frequency around 80 Hz. We calibrated each source time function prior to the inversion by estimating a matching filter between synthetic and field data using the limited offset range [$-250$ - $-200$ m]. We ran the inversions until reaching the convergence rate of 0.03\%. Figures \ref{fig:ch5un_fwi2d_2a} and \ref{fig:ch5un_fwi2d_3a_2} show the inverted perturbations for the conventional and extended FWI after 40 and 41 iterations, respectively. The conventional FWI led to unexpected high and noisy perturbations suggesting that the inversion is over-fitting the data and compensating for source footprints. The extended FWI yields a more stable result with apparent vertical jittering possibly due to the beating effect of guided waves. Figure \ref{fig:ch5un_fwi2d_3a_2}a shows a bulk increase in $V_{\mathrm{P}0}$ and a particular region with a larger perturbation indicated by the ellipse. Note that the inverted $V_{\mathrm{P}0}$ perturbation has a low vertical resolution so that localized anomalies may be smeared vertically by the inversion. The $V_{\mathrm{S}0}$ perturbation in Figure \ref{fig:ch5un_fwi2d_3a_2}b depicts a layered structure following that of the reservoir with a particularly higher velocity in the region around X=0.5 km.

We show in Figure \ref{fig:ch5un_profiles_2} the velocity profiles extracted from our 2D models along the well trajectory and compared them with the sonic logs. We calibrated the extracted $V_{\mathrm{P}}$ profiles by multiplying by $\sqrt{1+2\bar\epsilon}$, where $\bar\epsilon = 61\%$ is the average $\epsilon$ in the reservoir layer from our starting model. The sonic logs measure the effective velocity along the well but are sensitive to the short-scale vertical variations of the layered reservoir and the exact location of the well. Thus, a direct comparison with our 2D models is difficult since apparent variations could be due to either the lateral anomalies or the stratified structure of the reservoir. The strong anisotropy makes the $V_{\mathrm{P}}$ comparison even more difficult. Nevertheless, we observe that the conventional FWI model exhibits extreme and unlikely variations, whereas the extended FWI model follows the general trend of the sonic logs and remain stable. The trend also confirms the strong VTI anisotropy, which we estimated from the data and not by comparing vertical and horizontal well logs. The variations in the starting model are due solely to the layering effect. We observe little differences with the extended FWI mainly because the latter has a low vertical resolution in  $V_{\mathrm{P}0}$, so it is more relevant to assess it on the 2D perturbation in Figure \ref{fig:ch5un_fwi2d_3a_2}a. Interestingly, the $V_{\mathrm{S}}$ profile from the extended FWI seems to capture several variations observed in the well log and indicated by the black arrows in Figure \ref{fig:ch5un_profiles_2}b. These variations must correspond to lateral anomalies as highlighted by the ellipse in Figure \ref{fig:ch5un_fwi2d_3a_2}b and coincide with the $V_{\mathrm{P}0}$ anomaly in Figure \ref{fig:ch5un_fwi2d_3a_2}a. They may indicate a lower pore pressure or tighter shale region in the reservoir. Other velocity variations are opposite to the well log, in particular around $X=0.85$ km indicated by a red arrow in Figure \ref{fig:ch5un_profiles_2}b. However, since we see similar variations in the starting model, we conclude that such variations are due to the layering. Note that the $V_{\mathrm{S}}$ profiles are that of $V_{\mathrm{S}0}$, and that the SV-arrivals, which mostly constrain $V_{\mathrm{S}0}$ in the inversion, are not affected by the VTI anisotropy since the wave propagation is mainly horizontal.  

\plot[!h]{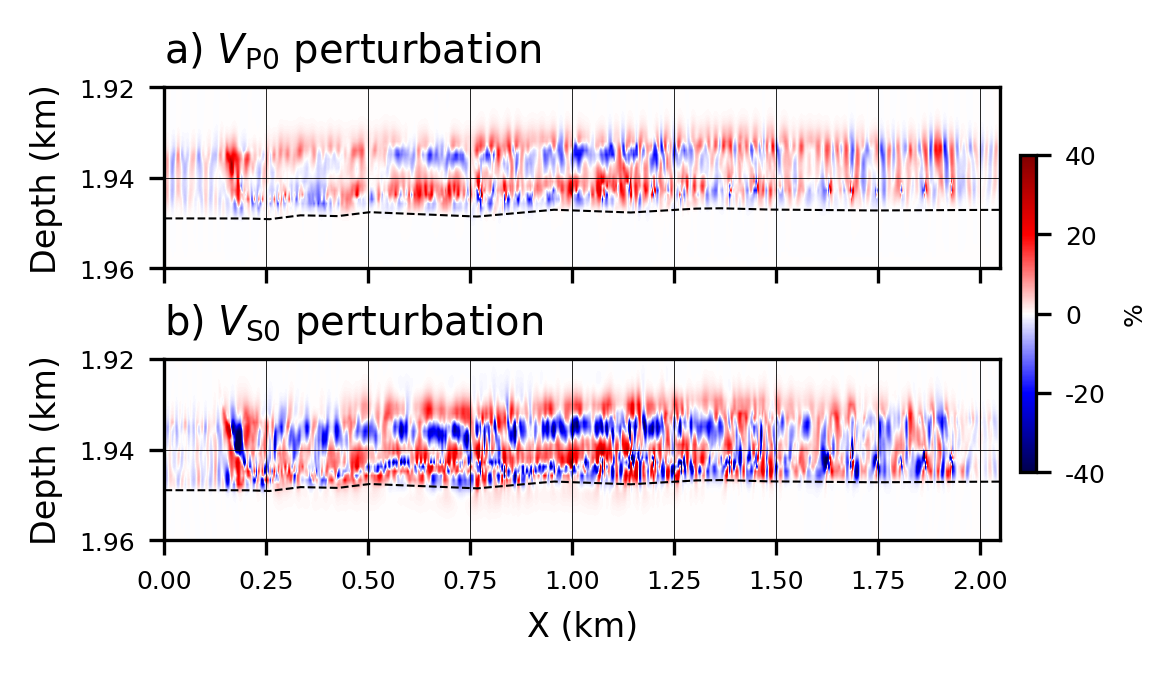}{width=0.95\textwidth}{Inverted perturbations using conventional FWI. The dashed line indicates the reservoir bottom horizon. The perturbations recovered upon inversion completion are unrealistically high and exhibit a strong noise caused by the presence of seismic sources inside the waveguide.}

\plot[!h]{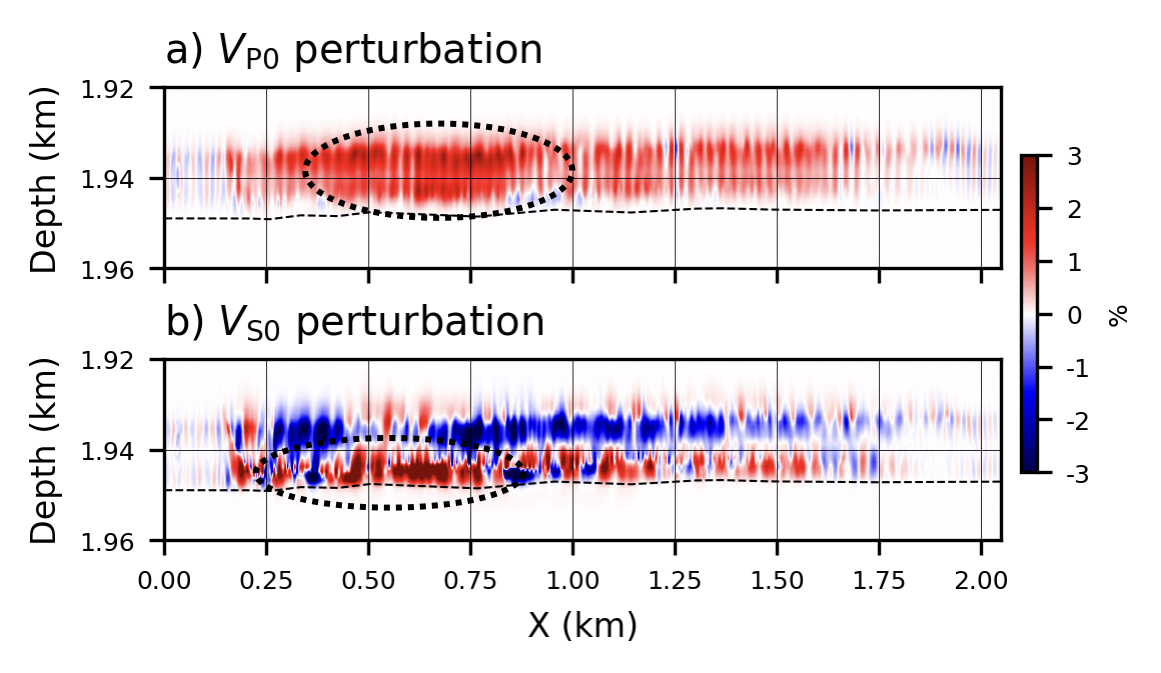}{width=0.95\textwidth}{Inverted perturbations using extended FWI. The perturbations are stable and moderate as expected given the lateral homogeneity of the unconventional reservoir before stimulation. A region of interest is delimited by the ellipsis, indicating a possible lower pore pressure or tighter shale part of the reservoir.}

\plot[!h]{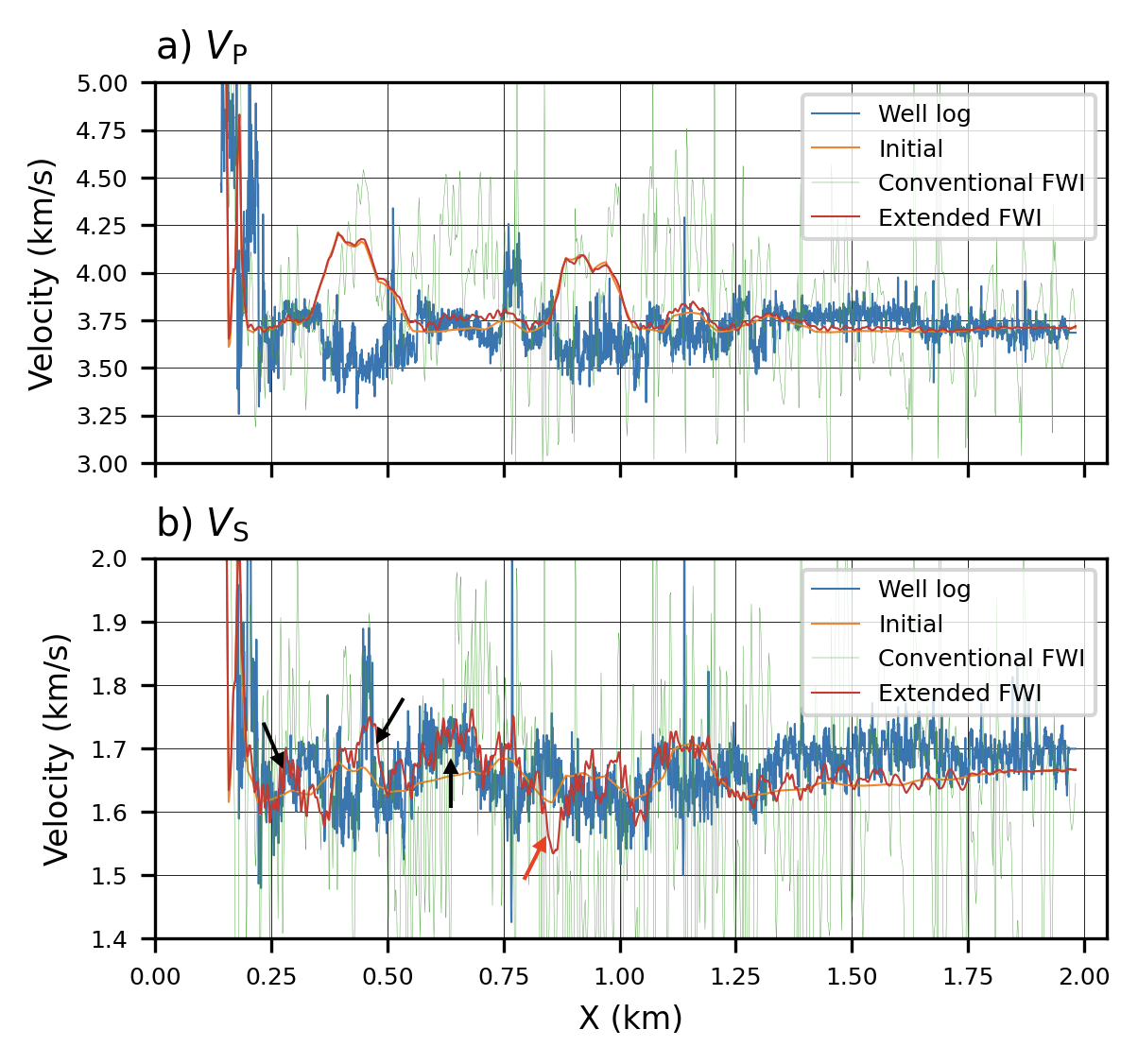}{width=0.95\textwidth}{Sonic logs and velocity profiles extracted from the 2D models along the well trajectory. The variations in the logs could be related to lateral anomalies and/or vertical layering. The conventional FWI exhibits wild variations whereas the extended FWI yields more stable result following the overall trend in the reservoir. Several variations in the $V_{\mathrm{S}}$ profile, indicated by the black arrows, could be localized anomalies and are corroborated by the variations observed in 2D in Figure \ref{fig:ch5un_fwi2d_3a_2}. Other variations, such as the one indicated by the red arrow, may be due to the layering effect.}

We further check the quality of the extended FWI model by comparing synthetic and field data. Figures \ref{fig:ch5un_model_2d_interleave_2} and \ref{fig:ch5un_fwi2d_3a_interleave_2} show three shot gathers distributed across the reservoir for the starting and extended FWI models, respectively. We interleave field and synthetic traces for an easier comparison. The synthetic traces are indicated by the blue boxes on top. We clip the amplitudes of all shot gathers to the same scale. The starting model already yields a good fit with no apparent cycle-skipping for P- and SV-arrivals in the inverted frequency band. However, there are subtle misalignment indicated by the arrows that the inverted model was able to correct.

\plot[!h]{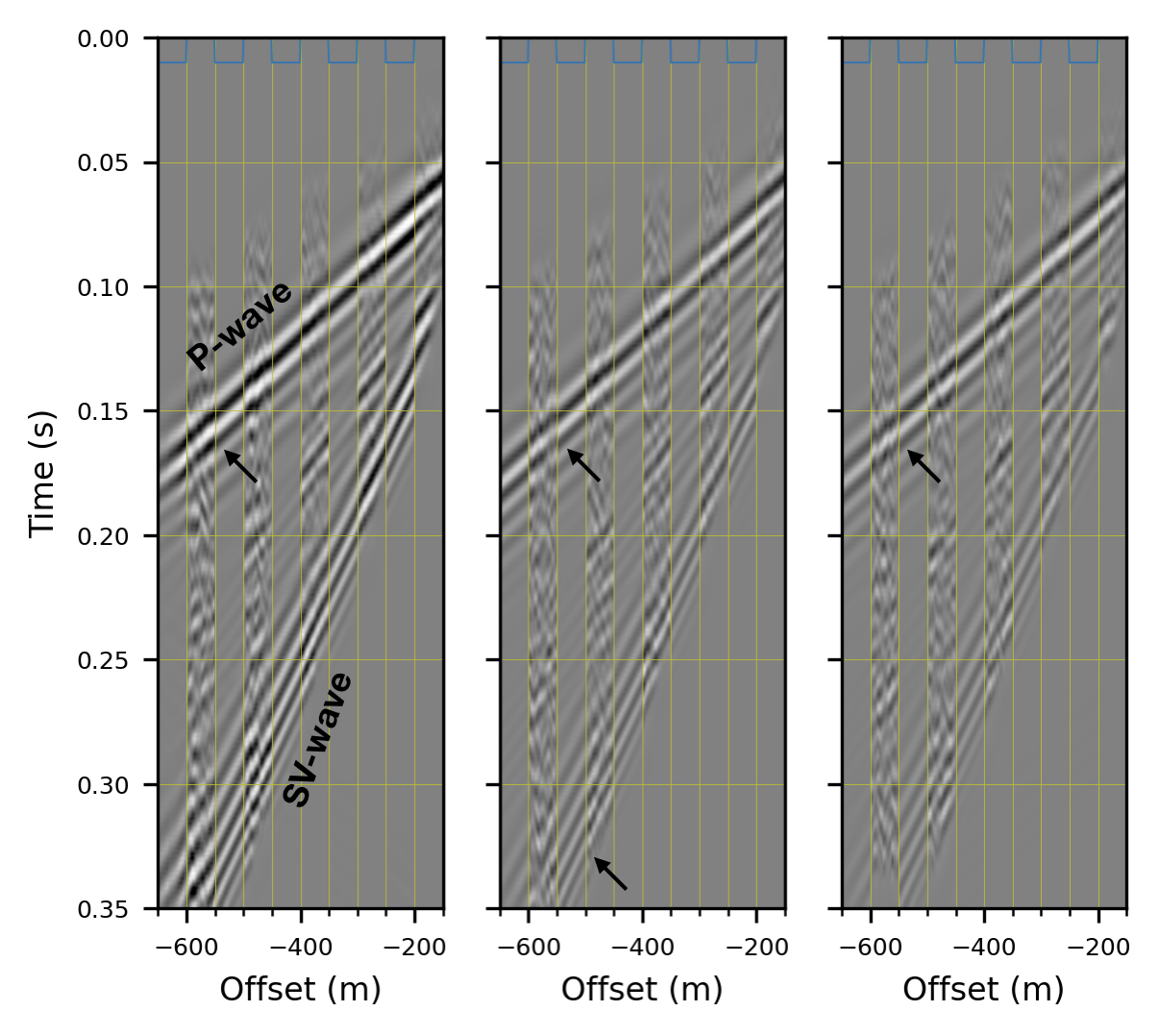}{width=0.95\textwidth}{Three shot gathers distributed across the reservoir showing interleaved field and synthetic traces. The latter are indicated by the blue boxes on top and are modeled using the starting 2D VTI model. The overall fit is good with no indication of cycle-skipping. However, subtle misalignment can be observed (arrows) which indicates model variations not accounted for.}

\plot[!h]{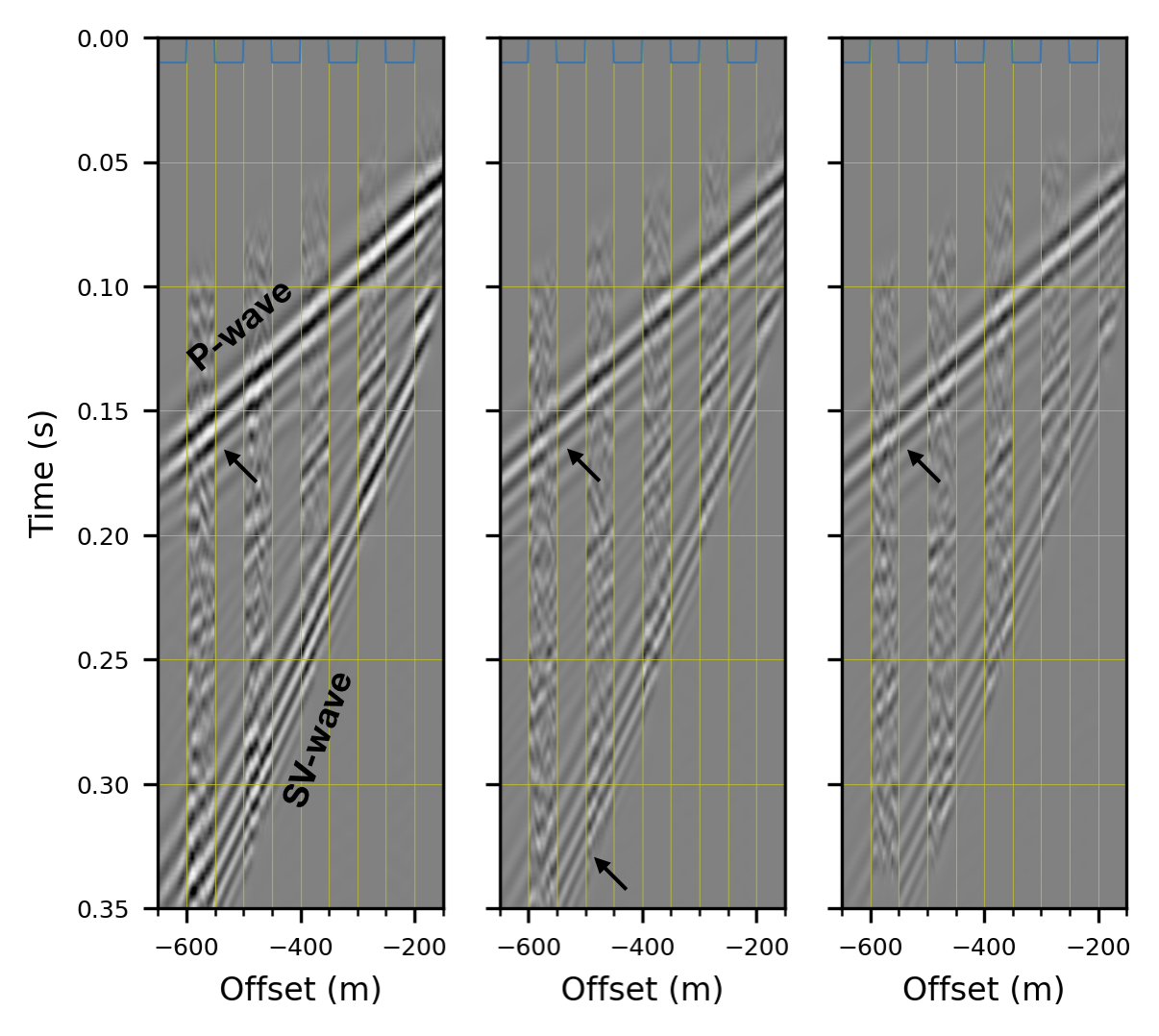}{width=0.95\textwidth}{Same shot gathers shown in Figure \ref{fig:ch5un_model_2d_interleave_2} where the synthetic traces are modeling using the extended FWI model. The inversion has improved the alignment with field data.}

To better understand why conventional FWI leads to too large perturbations upon inversion completion, we stopped it at iteration 21 where the data misfit is comparable to the extended FWI. Figure \ref{fig:ch5un_fwi2d_2b} shows the corresponding model perturbations. The latter depict some similarity with the perturbations from the extended FWI, in particular in the areas highlighted by ellipses in Figure \ref{fig:ch5un_fwi2d_3a_2}. However, the magnitude of the perturbations are still too high, and source footprints are prominent, especially in the $V_{\mathrm{S0}}$ perturbation. The subsequent iterations were over-fitting the data while compensating for the source footprints, which explains the wild variations in conventional FWI. We added a smoothing regularization to both inversions. Figures \ref{fig:ch5un_fwi2d_2c} and \ref{fig:ch5un_fwi2d_1} show the inverted perturbations for the regularized conventional and extended FWI after 15 and 64 iterations, respectively. We implemented the regularization by parameterizing the model in terms of cubic B-spline functions \cite[]{de1978practical,barnier2019waveform} with 50 m node spacing in the X-direction, and 1 m spacing in the vertical direction. Regularizing the conventional FWI smooths the source footprints without effectively removing them, leading to blobby perturbations. Conversely, further regularizing the extended FWI is able to smooth the vertical jittering without compromising the detected anomalies. Figure \ref{fig:ch5un_data_misfit} shows the normalized data misfit for conventional and extended FWI, with and without smoothing regularization. The extended FWI typically requires more iterations to reach convergence or a given data misfit. However, even at a data misfit similar to conventional FWI, the extended FWI yields more stable and realistic results. 

\plot[!h]{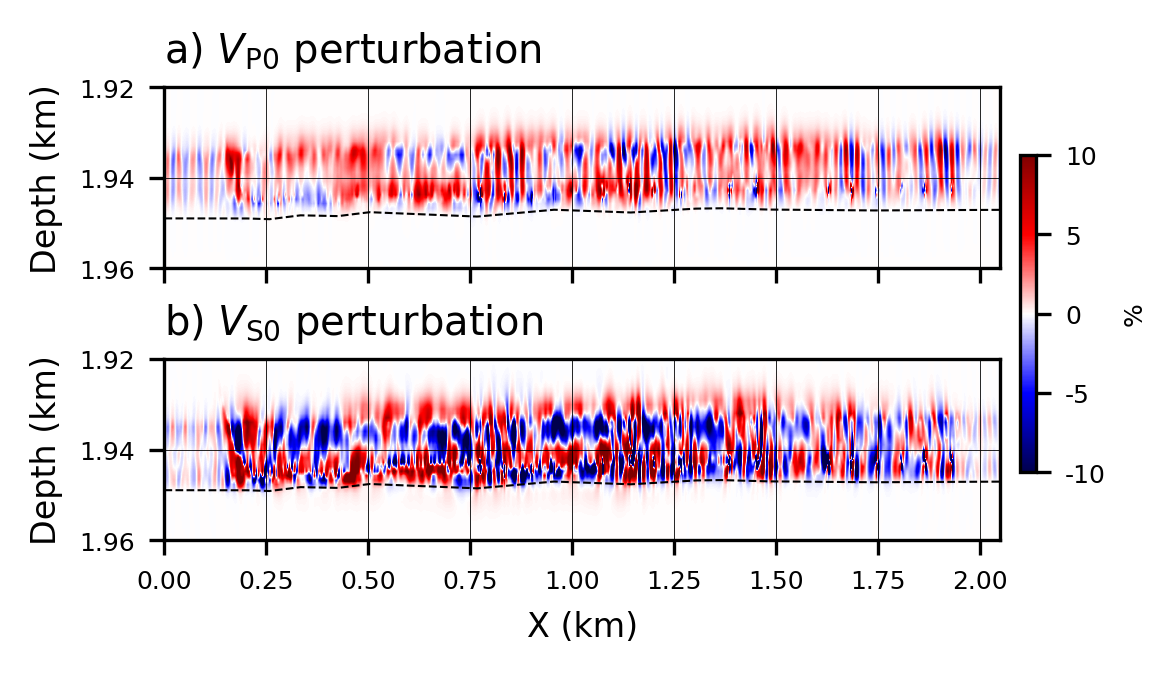}{width=0.95\textwidth}{Inverted perturbations using conventional FWI after 21 iterations only. The perturbations share some similarities with the extended FWI in Figure \ref{fig:ch5un_fwi2d_3a_2}. However, source footprints are clearly visible, and perturbations magnitude is too high.}

\plot[!h]{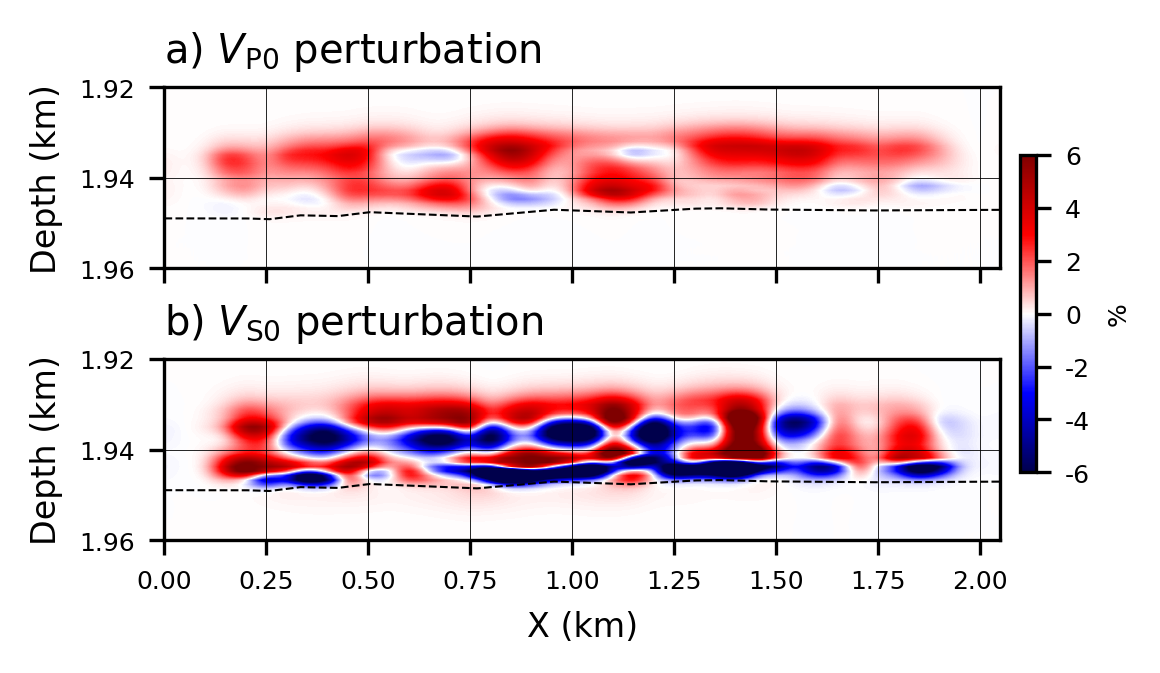}{width=0.95\textwidth}{Inverted perturbations using conventional FWI with B-spline model parameterization. The B-spline smoothing reduced the noise and smeared source footprints without removing them, which led to unlikely blobby perturbations.}

\plot[!h]{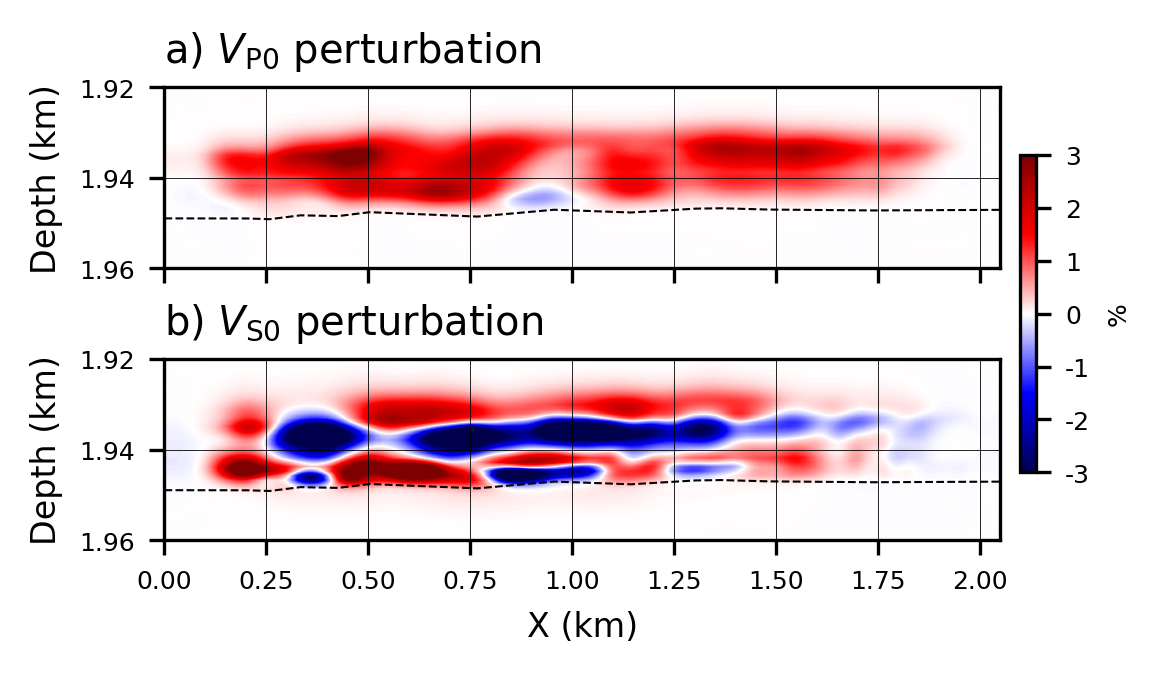}{width=0.95\textwidth}{Inverted perturbations using extended FWI with B-splines model parameterization. The B-splines play a regularization role and yield a smoother model that is consistent with the original extended inversion.}

\plot[!h]{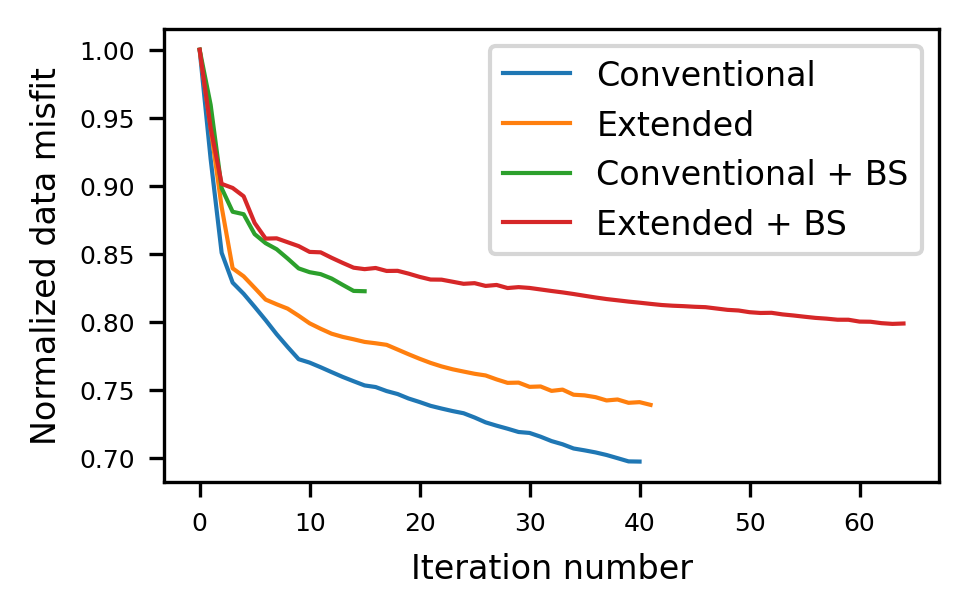}{width=0.95\textwidth}{Normalized data misfit for conventional and extended FWI, with and without B-spline (BS) parameterization.}

\section{Computational cost}

The computational cost of one iteration of extended FWI is almost the same as conventional FWI since the overhead caused by the additional operators in misfit \ref{misfit-extended} is negligible compared to wave propagation. However, the memory required to store the extended model and the corresponding weights increases linearly with the number of sources. For surveys with large number of shots, it may be more efficient to group close-by shots and extend the model along groups of sources instead of individual ones. Given the small number of shots in this work, we adopted a conservative approach of extending the model for each source. Moreover, we set the damping parameter $\eta$ appearing in the extended misfit \ref{misfit-extended} to a high-enough value to enforce consistency between the components of the extended model throughout the inversion. Strongly enforcing this regularization may require more iterations to reach a data misfit comparable to conventional FWI, which increases the overall cost of extended FWI.
\section{Conclusions}

Source footprints present a long-lasting challenge to FWI and can be detrimental when the sources or receivers are located inside the region of interest. Source-related errors such as time function, moment tensor, and location exacerbate these footprints and make the FWI models unusable. We have introduced a novel extended FWI formalism that uses illumination redundancy to eliminate the footprints and yield more robust inversions. The method outperforms other approaches, such as source-illumination compensation and gradient preconditioning in guided wave inversion settings. Although it requires more iterations to reach a specific data-fitting metric, the cost of each iteration is comparable to conventional FWI. Moreover, our method ensures more robust results.
We applied our method to the elastic FWI of a field DAS dataset. We showed that it could retrieve subtle anomalies near the sources in the unconventional reservoir, which are consistent with a horizontal well log traversing the reservoir. Such anomalies may indicate a lower pore pressure or a tighter shale formation. The conventional FWI yields unrealistic model perturbations for the same dataset. Regularizing the inverse problem to produce smoother models is effective only after removing the source footprints using extended FWI. Our method can also be applied to inversions with microseismic events to update the subsurface model in their vicinity. Other possible applications include vertical seismic profiling (VSP), where physical or virtual sources are located inside the well, surface-wave inversion to estimate near-surface models, and regional waveform inversion using earthquakes as sources.

\bibliographystyle{seg}


\end{document}